# A new method for interpreting well-to-well interference tests and quantifying the magnitude of production impact: Theory and applications in a multi-basin case study

*Mouin Almasoodi[1], Thad Andrews[1], Curtis Johnston[1], Ankush Singh[2], and Mark McClure[3]*

*[1]Devon Energy, [2]ResFrac Corporation*

## Abstract

Interference tests are used in shale reservoirs to evaluate the strength of connectivity between wells. The results inform engineering decisions about well spacing. In this paper, we propose a new procedure for interpreting interference tests. We fit the initial interference response with the solution to the 1D diffusivity equation at an offset observation point. It is advantageous to use the initial interference response, rather than the subsequent trend, because the initial response is less affected by nonlinearities, time-varying boundary conditions, and uncertainties about flow geometry and flow regime. From the curve fit, we estimate the hydraulic diffusivity and the conductivity of the fractures connecting the wells. For engineering purposes, it would be useful to quantify the impact of interference on well production. Thus, we seek a relationship between the 'degree of production interference' (DPI) and an appropriate dimensionless quantity that can be derived from the estimate of fracture conductivity. Using simulations run under a wide range of conditions, we find that the classical definition for dimensionless fracture conductivity does not achieve a consistent prediction of DPI. This occurs because the dimensionless fracture conductivity was derived assuming radial flow geometry, but the dominant flow geometry during shale production is linear. We also find that the CPG (Chow Pressure Group) metric does not yield consistently accurate predictions of DPI. As an alternative, we derive a dimensionless quantity similar to the classical 'dimensionless fracture conductivity,' but which is derived for linear, not radial, flow geometry. Using this approach, we calculate a 'dimensionless interference length' that collapses all cases onto a single curve that predicts DPI as a function of fracture conductivity, well spacing, and formation properties. We conclude by applying the new method to field cases from the Anadarko and Delaware Basins.

## 1. Introduction

Lateral and vertical well spacing decisions are critical for economic optimization in shale (Cao et al. 2017; Li et al., 2020; Miranda et al., 2022). Tighter well spacing reduces production and ROI per well but increases production and NPV per section of land (McClure et al., 2022a). Companies balance these considerations to determine well spacing, based on their commercial objectives and productions constraints.

In order to optimize, companies need to be able to predict production as a function of spacing. However, it is not easy to measure the size and shape of the drainage volume around a well. Proppant settling and localized screenout prevent proppant from being placed along the full length of hydraulic fractures (Almasoodi et al, 2020; Cipolla et al., 2022; McClure et al., 2022b). Diagnostics such as microseismic and offset fiber can measure the full wetted length of fractures, but diagnostics to measure propped length are not readily available. Even if the propped areas from adjacent wells overlap, there is not an easy way to estimate the conductivity of the flow pathway between the wells, nor the degree to which the wells will affect each other's production.

Interference tests fill a critical gap because they measure the degree of communication between neighboring wells. If we can use interference tests to measure the degree of connectivity between wells at different distances, we can map out the relationship between spacing and interference, providing key information needed for spacing optimization.

In a typical interference test, one or more wells are progressively put on production, and pressure is measured in one or more shut-in offset wells. After each well is put online, the pressure changes in the offsets well(s) are analyzed to assess the degree of connectivity.

The Chow Pressure Group (CPG) method is often used to interpret interference tests (Chu et al., 2020; Miranda et al., 2022). The method is based on fitting a power law curve to the pressure response at the monitoring well(s) after the offset well(s) are put on production. The power law exponent is used to calculate the 'CPG' parameter, which is related to the degree of connectivity between the wells.

A limitation of the CPG is that does not yield a quantitative estimate of how much production at each well is impacted by production of its neighbors. Further, it is not clear whether the same value of CPG measured between two wells always corresponds to the same amount of production interference, regardless of reservoir properties, fluid properties, or test conditions.

In this paper we present the Devon Quantification of Interference (DQI) procedure. The procedure interprets the interference test to estimate the hydraulic diffusivity and the fracture conductivity between the wells. A 'dimensionless fracture spacing' parameter ($L_D$) is calculated using the conductivity estimate and knowledge of reservoir parameters. Then, the Degree of Production Interference (DPI) is estimated from $L_D$. We performed simulations with different formation properties, fluid properties, well spacings, and conductivities, and found that the relationship between $L_D$ and DPI collapses onto a single curve.

We define a new metric, DPI, to quantify the effect of well-to-well interference on production. Consider the following scenario: two wells have been producing for at least several months. The two wells are bounded by each other on one side, and unbounded on the other side. Then, one of the wells is shut-in. What is the impact on the production rate of the well that remains on production? To quantify, we define the parameter 'degree of production interference' (DPI) as:

$$DPI = \frac{q_{20days} - q_{bef}}{q_{bef}} \quad (1)$$

where $q_{20days}$ is the production rate of the producing well 20 days after the shut-in of the offset well, and $q_{bef}$ is its production rate prior to the shut-in. If DPI is 1.0, this implies that the production rate of the active well doubles after the offset well is shut-in. If DPI is 0.1, this implies that the production increases by 10%.

The DPI value cannot exceed 1.0 because it is based on the hypothetical assumption that the two wells are unbounded on the outside, and from symmetry, the effect on production from the bounded side cannot exceed the production from the unbounded side. For more complex well configurations (such as wells bounded on both sides), it is hypothetically possible that production could more than double from the shut-in of the neighboring well. However, this would depend on details such as the spacing to the offset well on the other side, whether there is yet another well beyond the shut-in well, and whether there are additional wells above or below. For purposes of defining a metric, it is useful to avoid these complexities, and so we define the DPI to be strictly valid for the specific case that there are only two wells, and they are not bounded on the outside. This enables 'apples to apples' comparison between tests, even if they have different well configurations and reservoir properties.

Why don't operators perform interference tests by shutting in a well, and then measuring the impact on production rate at the offset wells? This would be possible in some cases, but usually, production rate measurements are not usually taken with sufficient frequency or precision (especially when allocating production between wells in the same pad) for this to be practical, and well shut-ins are usually not longer than a few days.

This paper focuses on shale applications. However, the method in this paper could be applied in other settings with fracture connections between wells, such as Enhanced Geothermal Systems.

## 2. Review of the existing CPG based methodology to quantify interference

The CPG method is the most widely-used method for interpreting interference tests in unconventional reservoirs. However, the CPG method has limitations: it does not account for variability of fluid and rock properties, and while it considers the shape of the pressure trend during interference, it does not account for the magnitude or timing.

In this section, the overall CPG methodology is explained. For simplicity, we assume that there are only two wells – a Producing Well and a Monitoring Well. The Producing Well is put on production first, while pressure is monitored at the Monitoring Well. Subsequently, the Monitoring Well would be put on production.

1. Initially the wells under consideration are shut-in. Subsequently, one of the interfering wells is put on production (POP).
2. A curve is fit to the Monitoring Well pressure trend prior to the Producing Well being put on production. This may be either a linear or power law function. Then, $\Delta P$ after POP is calculated as the difference between the measured pressure and the extrapolated curve.
3. The Bourdet derivative (also known as semilogarithmic derivative) $\Delta p'$ is computed as:

$$\Delta p' = \frac{d\Delta p}{d(\ln(t))} \qquad (2)$$

4. The CPG is then computed as:

$$CPG = \frac{\Delta p}{2\Delta p'} \qquad (3)$$

5. After the CPG stabilizes, the Magnitude of Pressure Interference (MPI) can be calculated as the mean of the stable CPG values. Often, it is difficult to obtain stable CPG values in real field data, which can lead to error in the estimate of MPI.

Chu et al. (2020) propose that CPG less than 0.5 corresponds to weak interference, 0.5-0.75 as moderate interference, and CPG greater than 0.75 corresponds to strong interference. Detailed explanation of the CPG methodology is provided by (Ballinger et al., 2022).

## 3. The DQI procedure for quantifying interference

We propose a new methodology – the Devon Quantification of Interference (DQI) procedure – to analyze interference tests. The steps of the procedure are listed below and then discussed in detail in the subsequent subsections:

1. Create log-log plots of $\Delta P$ and $t\frac{dp}{dt}$ versus time
2. Estimate the hydraulic diffusivity of the fracture connection between the wells
3. Estimate the fracture conductivity of the connection between the wells
4. Calculate the dimensionless interference length, $L_D$
5. Using the correlation in Figure 3 and the calculated $L_D$, estimate the degree of production interference (DPI)

### 3.1. Step 1: Creating $\Delta p$ and Bourdet derivative plots

Figure 1a shows pressure observations at a monitoring well in a field example in the Meramec formation located in the Anadarko basin of Oklahoma. The vertical black line shows the time when the offset well was put on production, which caused a downward pressure response in the monitoring well.

To perform an interpretation, we start by constructing a log-log pressure derivative plot showing $\Delta p$ and $t\frac{dp}{dt}$ versus time from the beginning of the transient (Figure 1b). The value $\Delta p$ is defined as: $\Delta p = p - p_{ref}$. $p$ is the pressure recorded by the pressure gauge, and $p_{ref}$ is the extrapolation of the prior pressure decline trend (the orange line in Figure 1a). It may be useful to smooth the data by resampling at increments of pressure.

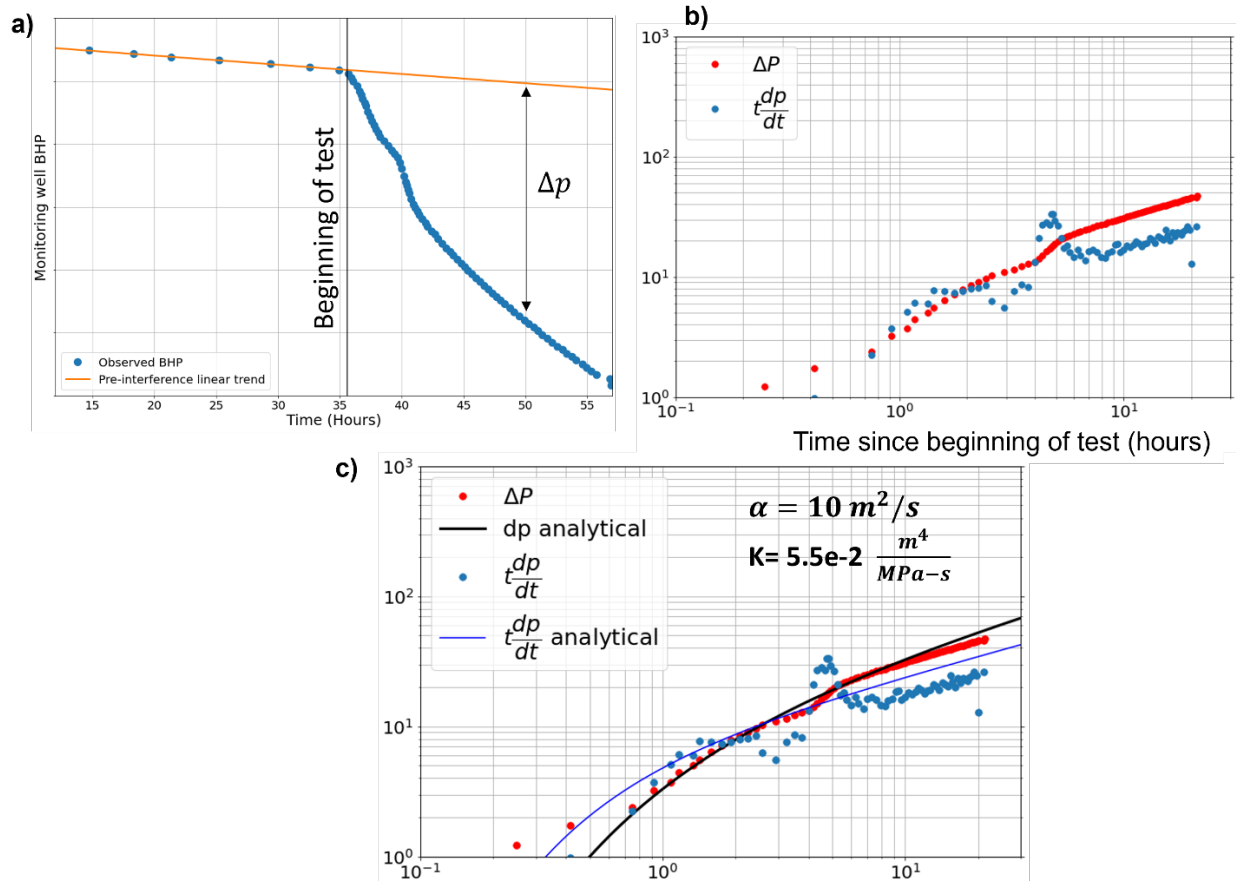

*Figure 1. Example of the plots required for interpreting an interference test using the DQI method. a) Pressure observations in a monitoring well before and after an offset well is put on production. The timing of the offset well POP is marked by the vertical black line, indicating the beginning of the interference test. The orange line shows the linear trend of the pressure prior to interference which is extrapolated to the interference test time period to calculate Δp; b) Δp and $t\frac{dp}{dt}$ observed during the interference test; c) Δp and $t\frac{dp}{dt}$ from the analytical solution (black and blue lines) are fitted to the pressure and derivative observations by a trial and approach varying α and K, to find a good fit.*

### 3.2. Step 2: Computing diffusivity from pressure changes in interference

When the active well is put-on-production, a pressure disturbance propagates along the fracture, as described by the pressure diffusion equation (Zimmerman, 2018) . In classical well test analysis, the 'radius of investigation,' $r_{inv}$, is defined as the distance from the well where a pressure disturbance will be felt, as a function of time, *t*, and hydraulic diffusivity, $\alpha$ (Horne, 1995):

$$r_{inv} = 4\sqrt{\alpha t} \tag{4}$$

For flow through porous media, the hydraulic diffusivity is equal to the square root of permeability divided by porosity, total compressibility, and viscosity. In a fracture, it is more appropriate to define hydraulic diffusivity in terms of the fracture conductivity and aperture:

$$\alpha = \frac{k_f W}{\mu \left(\frac{dW}{dp} + c_f W\right)} = \frac{C}{\mu \left(\frac{dW}{dp} + c_f W\right)} \tag{5}$$

where $k_f$ is the fracture permeability, $W$ is the aperture, $C$ is the fracture conductivity, $\mu$ is fluid viscosity, $\frac{dW}{dp}$ is the derivative of aperture with respect to pressure, and $c_f$ is the compressibility of the fluid in the fracture.

If we assume that the fracture is fixed-height, and that the pressure disturbance propagates along the fracture through linear flow, we can calculate the pressure change from the solution for a constant flux source in a semi-infinite slab (Hetnarski et al., 2014), with the pressure disturbance observed at an offset distance $y$.

$$P(y,t) - P(y,0) = \frac{2q_0 \sqrt{\frac{\alpha t}{\pi}}}{K} \exp\left(-\frac{y^2}{4\alpha t}\right) - \frac{q_0 y}{K} \operatorname{erfc}\left(\frac{y}{2\sqrt{\alpha t}}\right) \tag{6}$$

$q_0$ = production rate (m³/s)

$y$ = offset distance (m)

$\alpha$ = diffusivity (m²/s)

$K$: equal to $\frac{C*H}{\mu}$, where $C$ is fracture conductivity, $H$ is the fracture height, $\mu$ is the viscosity of the fluid inside the fracture (m⁴/MPa*s)

Plotting the analytical solution, we observe that the timing of the onset of the pressure interference is controlled by the diffusivity, and the shape of the curve after the onset of interference is controlled primarily by the $K$ parameter.

It is possible to fit measured data with Equation 6 and estimate both diffusivity and the $K$ parameter. However, the estimate of $K$ is considerably more uncertain than the estimate of diffusivity. It could be affected by: (a) non-constant production rate at the active well, (b) complex fracture geometries other than linear geometry, and (c) change in flow regime from fracture linear (for example, if flow from the matrix causes bilinear flow). Also, nonlinearities could be present during early-time production that – strictly speaking – may violate the linearized 'single phase' assumptions that justify: (a) the use a simple diffusivity equation solution (Equation 6), and (b) the calculation of $\Delta p = p - p_{ref}$ that subtracts out the prior pressure trend (which relies on superposition).

On the other hand, the estimate of diffusivity (from the timing of the onset of the signal) is quite robust. As shown in Equation 4, the radius of investigation scales with the square root of the product diffusivity and time, *regardless of flow regime, flow geometry, or variable boundary condition at the active well*. Also, the early-time pressure response of the interference test involves relatively small pressure changes, minimizing the potential impact of nonlinearities in the system, such as changing fluid compressibility.

Thus, the early interference pressure response – *the tip of the spear* – is the most robust part of the transient for estimating the diffusivity. It is least affected by uncertainties and nonlinearities.

To estimate diffusivity, we used a trial-and-error approach to fit Equation 6 to the observed $\Delta p$ and $\Delta p'$ curves. The value of *K* from the curve fit is not used in the rest of the analysis, but it is nonetheless useful to include in the curve fit that it is used for the estimation of the diffusivity. The priority is to fit the part of the trend that matches the initial response – the first 10s or 100s of psi. The analytical solution is not expected to be able to match the transient after the initial response. Figure 1c shows an example of fitting the analytical solution to a real interference test. Figure 2 shows sensitivities on the effect of changing the hydraulic diffusivity and *K*.

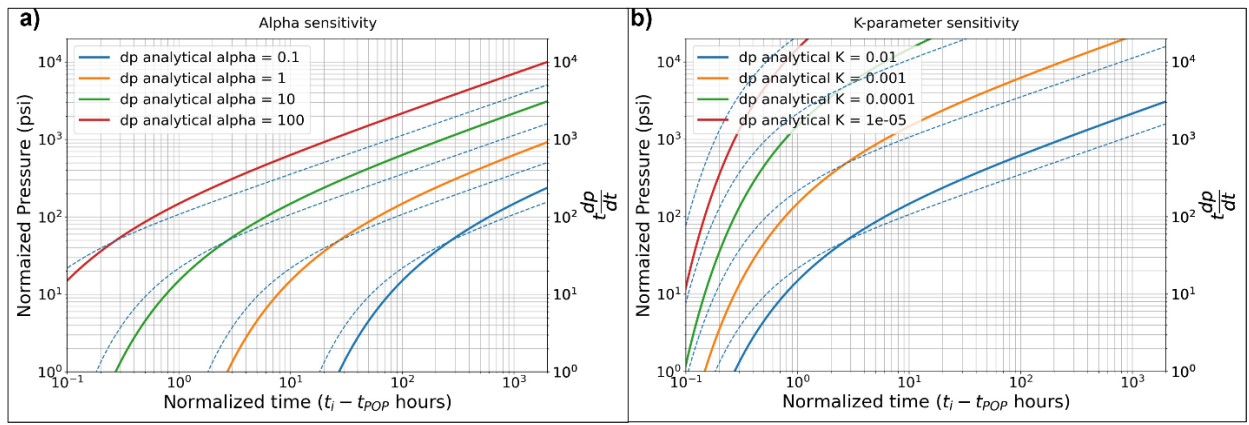

Figure 2: The variation in dp (solid lines) and $t\frac{dp}{dt}$ (dashed lines) at the offset Monitoring Well calculated with the analytical solution with different values of a) $\alpha$ (keeping K constant = 0.01), and b) K-parameter (keeping alpha constant = 10). The units of $\alpha$ are $\frac{m^2}{s}$ and the units of the K-parameter are $\frac{m^4}{MPa*s}$.

### 3.3. Step 3: Computing conductivity from diffusivity

Once the diffusivity has been estimated from the curve fit, we can plug into Equation 5 to estimate the fracture conductivity. We will not practically have precise knowledge of $\frac{dW}{dp}$ and $W$, but we can plug-in reasonable values: $\frac{dw}{dp} = 0.8e-5 \ to \ 3e-5 \frac{m}{MPa}$ ($2.2e-6$ to $8.1 e-6 \frac{in}{psi}$) and $W = 0.76 \ mm$ (0.03 in).

For $c_f$, we use the compressibility of the interfering fluid present in the fracture. For the initial POP test after stimulation, this will be the compressibility of water, $\sim 0.000435 \ MPa^{-1}$. For tests performed after months of production, the fluid in the fracture will be a multiphase mixture of oil, gas, and/or water. In this case, we must use the total compressibility of all the phases in the fracture (which can be estimated using the technique given in Appendix B).

For viscosity, we use the viscosity of the fluid in the fracture. The initial POP test may be controlled by the viscosity of the frac fluid. For instance, when a viscous HVFR is used for the hydraulic fracturing, the POP test viscosity may be influenced by the viscosity of the HVFR which can be significantly higher than water. During the later production phase tests, the effective viscosity of the multiphase mixture must be estimated (Appendix B).

### 3.4. Step 4: Calculating the dimensionless well spacing

Next, we calculate a dimensionless number, $L_D$, that can be roughly defined as the 'dimensionless drainage distance.' $L_D$ is derived by first approximating the 'maximum possible drainage distance along the fracture if hypothetically it was unbounded and had infinite propped length.' This drainage distance is determined by a balance of flow rate into the fracture and flow rate along the fracture.

The derivation of $L_D$ uses coarse 'back of the envelope' approximations, but this is acceptable because our objective is not to derive a rigorous analytical solution. Instead, we are seeking to determine an appropriate scaling between variables. Then, in Section 4.2, simulations are used to empirically verify that the postulated scaling is correct, and to determine the quantitative relationship between $L_D$ and DPI.

To estimate the flow rate along the fracture (towards the well), we write Darcy's law:

$$Q_{frac} = \frac{C_{fracture}}{\mu} H \frac{\Delta p}{L} \tag{7}$$

where

$Q =$ flow rate (m³/s)

$H =$ fracture height (m)

$\Delta p \approx BHP_{initial} - P_{res}$

$dL =$ Length of the drainage distance along the fracture (m)

Assuming matrix linear flow into the fracture, the fluid flow rate into the fracture can be expressed as:

$$Q = 2HL * (\Delta p) \sqrt{\frac{\phi c_t k}{\pi \mu}} t^{-0.5} \tag{8}$$

where

$\phi =$ porosity (v/v)

$c_t =$ total compressibility, i.e. fluid plus pore compressibility (MPa^-1)

$k =$ matrix permeability ($m^2$)

$t$ = time at which production impact is estimated (s) (used as 20 days for our analysis)

$L$ = Length of the region along the fracture where production is occurring

Next, we estimate the maximum possible length of fracture that could be drained, under the hypothetical assumption of unlimited propped length. Under these conditions, the draining length will be limited by the ability of the fracture to deliver fluid to the well, given the rate that fluid flows into the fracture.

The maximum possible draining length is estimated by setting the values of *Q* from Equations 7 and 8 to be equal (implying that the flow rate into the fracture is equal to the flow rate along the fracture) and solving for *L*. The flowing fluid density may be a bit different between the fracture and the matrix, and so it is not strictly precise to set the volumetric flow rates (*Q)* equal, but this is a minor approximation that simplifies the calculations. Further, we assume that $\Delta p$ for 'flow through the fracture' is approximately equal to $\Delta p$ for 'flow through the matrix,' and so the terms cancel-out (discussed further in Section 3.6 below).

$$L = \sqrt{\frac{(C_{fracture})\left(\frac{k_r}{\mu}\right)_{frac,t}}{2\sqrt{\frac{\phi c_t k}{\pi}\left(\frac{k_r}{\mu}\right)_{mat,t}} t^{-0.5}}} \tag{9}$$

$\left(\frac{k_r}{\mu}\right)_{frac,t}$ = the total mobility of the fluid in the fracture during production

$\left(\frac{k_r}{\mu}\right)_{mat,t}$ = the total mobility of the fluid in the matrix during production

Where total mobility is equal to $\frac{k_{rw}}{\mu_w} + \frac{k_{ro}}{\mu_o} + \frac{k_{rg}}{\mu_g}$

In these equations, because they are defining a production response during long-term depletion, the fluid properties should be defined for the flowing reservoir fluid, and not the properties of the frac fluid (even if the interference test is performed when the wells are initial put on production).

To build intuition, consider a few 'end-member' extremes. If the permeability is high relative to the conductivity, then fluid will be able to rapidly flow into the fracture, but the fracture conductivity (the ability to transmit fluid along the fracture) will limit the total flow rate, and so the effective draining length will be limited. Conversely, if the permeability is very low relative to the conductivity, then it will require a large producing fracture area to reach the flow capacity of the fracture, and the effective draining length will be large.

The balance of these two effects – the ability of fluid to transmit along the fracture and the rate that fluid can enter the fracture – determines the length of the fracture that can be drained. Of course, in reality, the propped length is not infinite. If the wells are so far-apart that they do not have overlapping propped areas, then the interference test results will be straightforward to interpret – there will be minimal interference. But, if the propped areas *do* overlap, then interference will occur, and Equations 7 and 8 can help quantify the magnitude of interference.

We next calculate the dimensionless drainage length, $L_D$, by dividing by the half-well spacing. We intuitively expect that DPI should increase with larger values of $L_D$.

$$L_D = \frac{L}{\frac{y}{2}} \tag{10}$$

$y$: well spacing

### 3.5. Step 5: Relating the dimensionless well spacing to the DPI

As discussed in Section 4.2, we ran numerical simulations of interference tests under a wide range of conditions, varying fluid properties, reservoir properties, and well spacing. In all cases, we simulated both: (a) an interference test, and (b) the production impact of shutting-in one of the wells. The interference test was analyzed to infer $L_D$, and the simulation of the single-well shut-in allowed us to directly observe the DPI. Then, we made a cross-plot of DPI versus $L_D$. Figure 3 shows that the results collapse onto a single curve. This means that we can use an estimate of $L_D$ to estimate the DPI.

After estimating $L_D$ from the interference test, the value of DPI can be read directly from Figure 3, or calculated from the equation:

$$DPI = \frac{1}{1 + (2.5/L_D)^{3.1}} \tag{11}$$

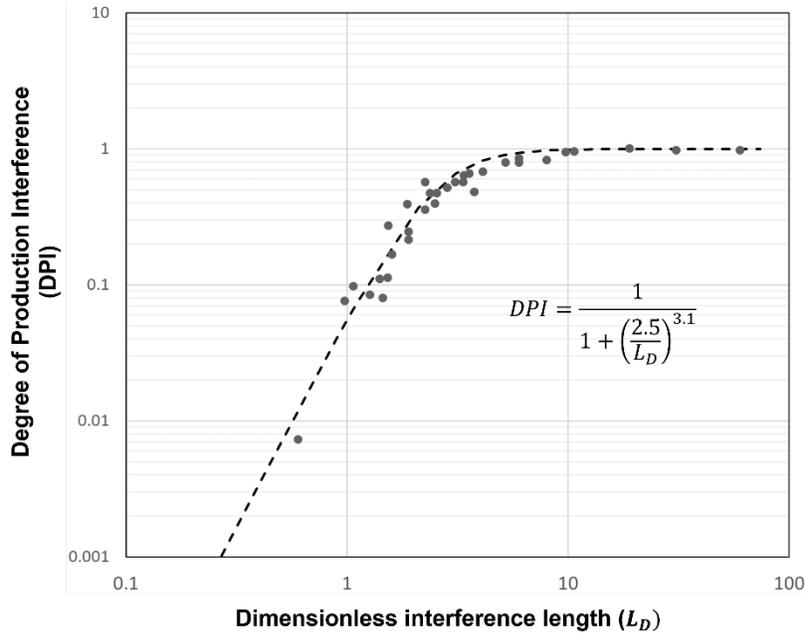

*Figure 3. The relationship between $L_D$ and DPI from numerical simulations collapse onto a curve represented by the dashed line. The dots represent the individual numerical simulations, which are described in Section 4. The simulations have varying fracture, fluid and matrix property inputs.*

### 3.6. Understanding why the classical 'dimensionless fracture conductivity' does not accurately capture the scaling between the variables

Conventionally, the concept of 'dimensionless fracture conductivity' ($F_{CD}$) (e.g., Pearson, 2001) has been used to estimate the relative strength of 'pressure drop along the fracture' and 'pressure drop in the reservoir.' In Section 3.4, the quantity $L_D$ is derived in an equivalent way – by balancing pressure drop along the fracture with pressure drop within the matrix. Instead of using $L_D$, why don't we use $F_{CD}$ to scale fracture conductivity, and seek a relationship between $F_{CD}$ and DPI?

Using the same set of simulations as discussed in Section 4.2, we plotted DPI versus $F_{CD}$, we found that the simulations *did not* collapse onto a curve, which suggests that $F_{CD}$ does not capture the scaling between relevant variables.

The reason is that the dimensionless fracture conductivity is derived from a ratio of pressure drops *specific to radial and boundary dominated radial flow* (Cinco-Ley. et al., 1978; Abbaszadeh & Cinco-Ley, 1995). This was an appropriate assumption in classical reservoir engineering in conventional reservoirs. However, in shale, fractures never enter radial flow, and instead experience a linear flow geometry perpendicular to the fracture faces.

The scaling between variables during radial and linear flows are different. For example, radial flow scales inversely with reservoir viscosity, but linear flow scales inversely with the square root of reservoir viscosity. In comparison, pressure drop along the fracture scales inversely with viscosity. Thus, if we take the ratio of 'radial flow' pressure drop and 'along the fracture' pressure drop, the viscosity terms will cancel – which is why viscosity does not need to be considered in the definition of $F_{CD}$. But, with linear flow geometry into the fracture, if we take the ratio of 'matrix linear flow' pressure drop with 'along the fracture' pressure drop, the viscosity terms *do not* cancel, which is why viscosity remains a term in Equation 9. Thus, for linear flow, the classical definition of $F_{CD}$ does not fully capture the scaling between key variables.

Equations 7 and 8 can be used to define a 'linear-flow equivalent' dimensionless fracture conductivity. We set Q through the fracture and matrix equal, and solve for the ratio of $\Delta p$ in the matrix and $\Delta p$ in the fracture:

$$\frac{\Delta p_{matrix}}{\Delta p_{fracture}} \propto F_{CD,linear} = \frac{C\left(\frac{k_r}{\mu}\right)_{frac,t}}{2L^2 \sqrt{\frac{\phi c_t k}{\pi}\left(\frac{k_r}{\mu}\right)_{mat,t}} t^{-0.5}} \tag{12}$$

This definition for $F_{CD,linear}$ is provided only for discussion purposes. The DQI correlation uses $L_D$, rather than $F_{CD,linear}$.

In Appendix D, we use a similar approach to derive the classical definition of $F_{CD}$, assuming radial flow, and show why it yields a different scaling than the derivation with linear flow.

## 4. Validation using simulation examples

We validated our proposed methodology and compared the results to the CPG workflow for both 3D numerical simulations and field data.

### 4.1. Demonstration of fracture conductivity estimates using constant rate simulations

#### 4.1.1. Simulation set-up

Utilizing a fully coupled 3D hydraulic fracturing, geomechanics, and reservoir simulator (McClure et al., 2022c), we performed a series of simulations of interference between hydraulically fractured wells.

In the base simulation, we used the simplest possible problem setup to isolate the basic properties of the system. Subsequently, we ran more complex simulations to assess the robustness of the interpretation procedure.

Figure 4 shows the model setup. It consists of two wells, the Production Well and the Monitoring Well, connected by a rectangular, constant conductivity fracture (Figure 4b & 2c). For simplicity, the process of hydraulic fracturing and proppant placement is not modeled; instead, the crack is specified as a 'preexisting' fracture with known properties. Simulations including full hydraulic fracturing/reservoir simulation are described in Section 4.3.

The initial aperture of the fracture is set to reasonable, generic values. The fracture aperture is calculated as a function of effective normal stress using the Barton-Bandis equation (Barton et al., 1985; Willis-Richards et al., 1996). The parameters of the Barton-Bandis equation are chosen such that the fracture compressibility is nearly constant under the range of conditions of the simulation and reasonable for a proppant pack conductivity (~0.01 $MPa^{-1}$). In each simulation, the fracture conductivity is set to a constant, uniform value.

The baseline simulation uses a single phase slightly compressible fluid. The permeability is homogeneous, and the pore pressure is hydrostatic. The water viscosity is 0.31 cP, and the fluid compressibility is $0.000435\ MPa^{-1}$.

Both wells are shut-in at the beginning of the simulation. The Production well is put-on-production (POP'ed) with a constant liquid production rate boundary condition of 2 bbl/day, as shown by the dashed blue line in Figure 4a. After the well is put on production, the Monitoring Well bottomhole pressure begins to decline.

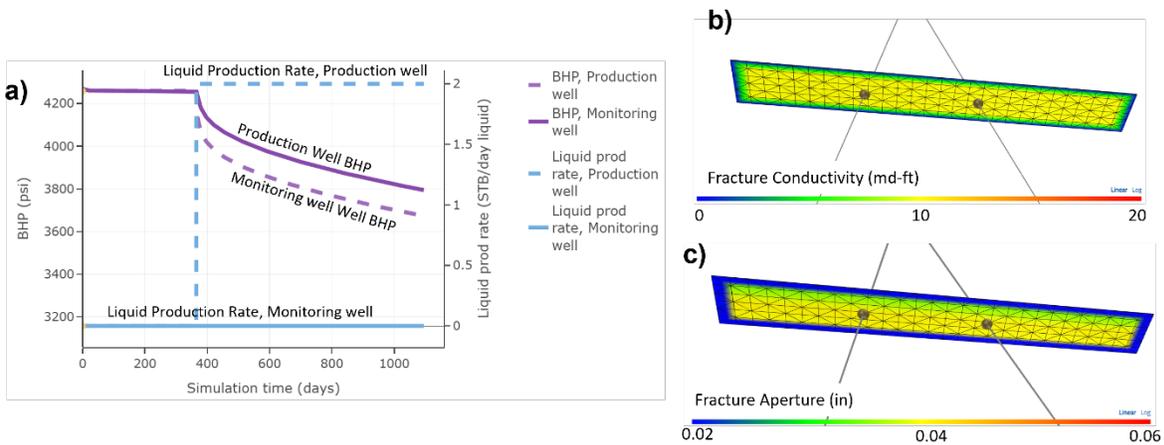

*Figure 4: a) Setup for the simple cases with a single interference test; B) Constant conductivity fracture connecting the two wells; C) Nearly uniform fracture aperture. Note that the conductivity and aperture appear to taper to zero at the edges of the fractures; however, this is a visualization artifact related to color interpolation. In the actual simulation, conductivity and aperture are nearly uniform.*

The initial baseline simulations were run with fracture conductivities of 0.1, 1, 10 and 100 mD-ft. Based on the simulation results, the diffusivity was computed using the procedure described in Section 3.

### 4.1.2. Results

Figure 5 shows that the analytical solution fits the early part of the pressure decline in the Monitoring Well and then deviates after a few weeks. The initial pressure response at the Monitoring Well, the 'tip of the spear,' represents fracture linear flow. The signal at later times is affected by transitions to bilinear flow and/or matrix linear flow. This is why the focus in the DQI methodology is to match the pressure decline in early period immediately after the onset of the signal. This avoids the complexities caused by the flow regime transitions and other nonlinearities.

As expected, the inferred $\alpha$ in the four cases increases with fracture conductivity. For each case we compute a range of possible fracture conductivities to account for the range in the assumed reasonable values of $\frac{dW}{dp}$ (as listed in Section 3.3). Table 1 shows the computed fracture conductivity ranges in the different cases and shows that the method can be used to get reasonable estimates which are consistent with the input fracture conductivity.

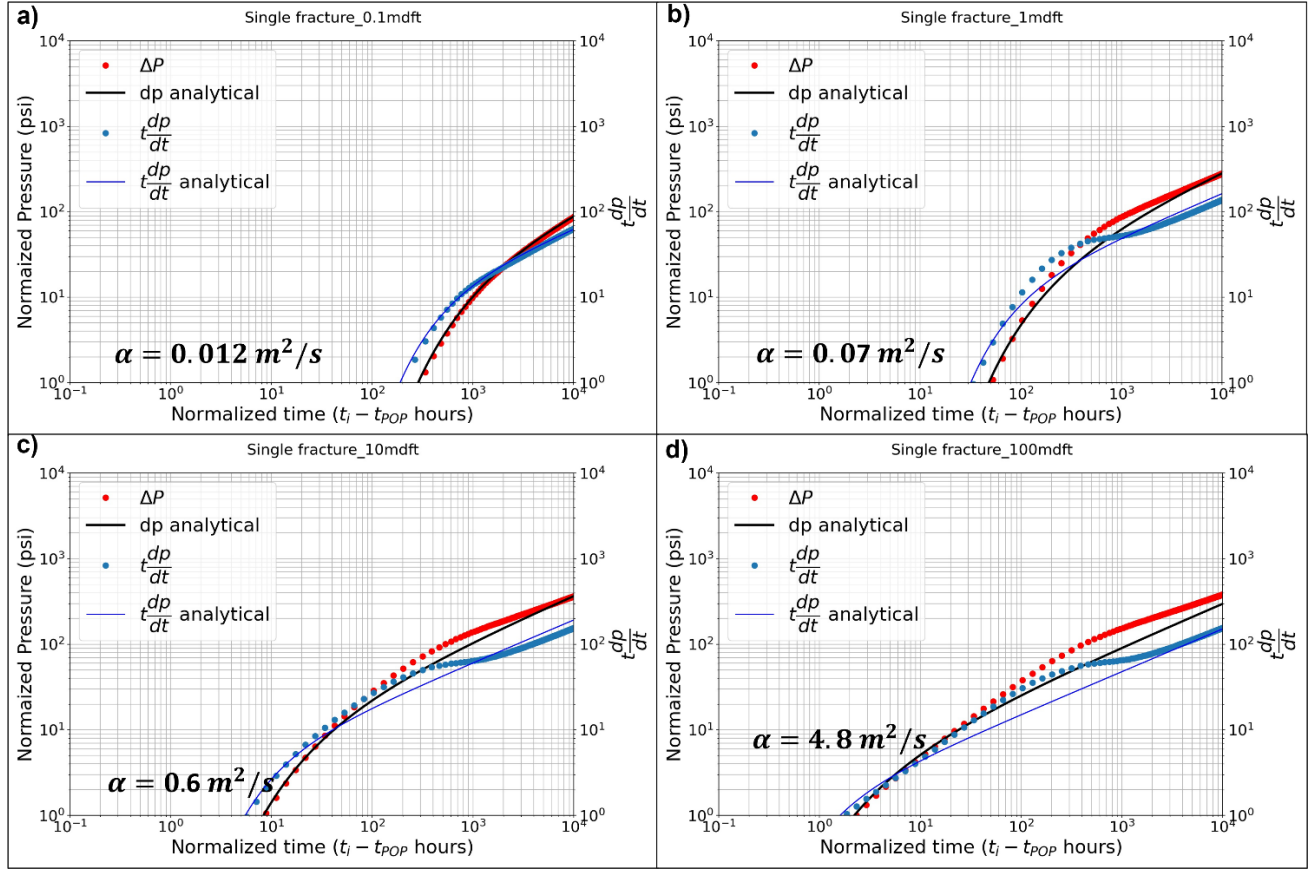

Figure 5: Constant rate cases showing the diffusivity analysis to calculate the conductivity of the connected fracture using the fracture linear flow analytical solution. The computed $\alpha$ is proportional to the input fracture conductivity.

Table 1: Fracture conductivity calculations from estimates of the hydraulic diffusivity

| Input fracture conductivity | Fitting parameters | | Estimated fracture conductivity | |
|---|---|---|---|---|
| | $\alpha$ | $K$ | Low | High |
| mD-ft | m2/s | $\dfrac{m^4}{MPa*s}$ | mD-ft | mD-ft |
| 0.1 | 0.012 | 2.50E-05 | 0.1 | 0.2 |
| 1 | 0.07 | 2.00E-05 | 0.6 | 1.4 |
| 10 | 0.6 | 5.00E-05 | 5.1 | 12.4 |
| 100 | 4.8 | 1.80E-04 | 40.7 | 99.3 |

### 4.1.3. Testing the analysis procedure under more complex and realistic conditions

To test the DQI method, we performed simulations with different conductivities, well spacings, viscosities, fluid compressibilities, and matrix permeabilities. In these simulations, instead of using a constant production rate boundary condition, we varied the BHP at the Production Well with a simplified power-law decline (based on the field dataset in Section 5.1).

The baseline simulation was extended to model multiple interference tests as shown in Figure 6. The initial interference test at POP – Test-1 – is followed by one year of production, after which the Monitoring Well is shut-in. The change in the production rate at the Production Well – after the Monitoring Well is shut-in – is used to quantify the degree of production interference between the wells (DPI). The Production Well is then shut-in and the change in pressure build-up of the Monitoring Well is observed in Test-2. Finally, the Production Well is put online and the pressure change in the Monitoring Well is observed. This sequence of events provides three different interference tests and also provides a direct measurement of 'production uplift' at the Production Well after the shut-in of the Monitoring Well. The fracture conductivity is kept the same throughout the simulations.

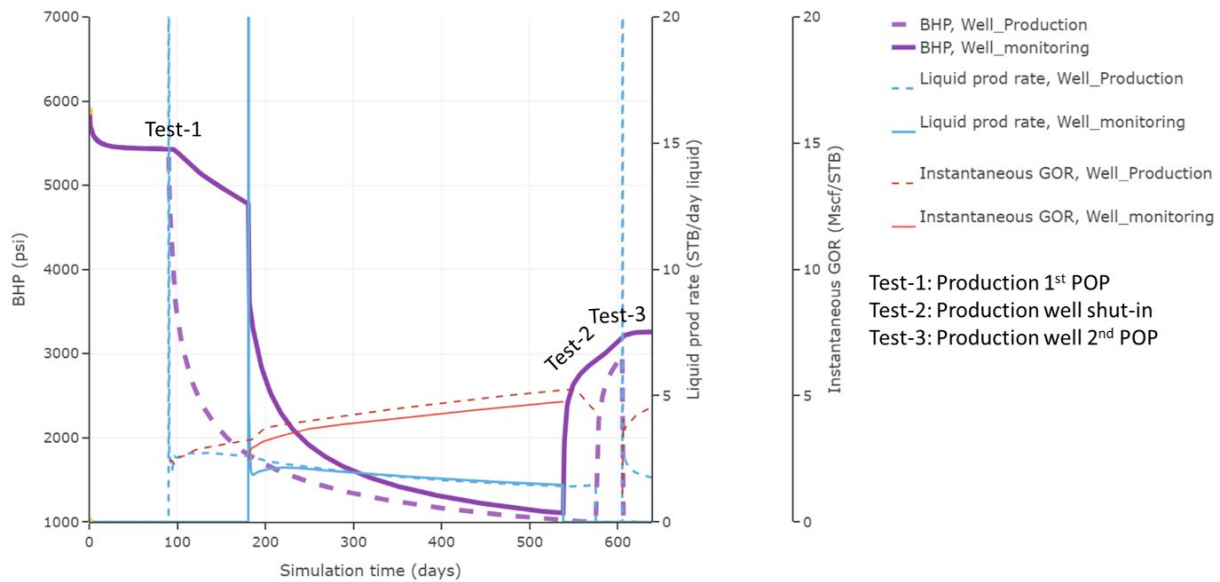

*Figure 6 Simulation setup with multiple tests and BHP pressure control with a realistic pressure decline to enable quantification of production rate change with interference.*

We ran the following simulations:

1. 10 mD-ft fracture conductivity with 1, 10, 20, 50, 100, 200, 500 nD matrix permeability
2. 1, 100 mD-ft fracture conductivity with 1, 10 and 100 nD matrix permeability
3. Formation fluid with viscosity 30x higher and 30x lower than water viscosity
4. Formation fluid with 10x and 100x higher compressibility than water
5. Different well spacing: 520 ft, 600 ft, 1140 ft, 1450 ft and 1760 ft

Figure 7 shows examples from four of the simulations. The early part of the BHP from the Monitoring Well fits well with the analytical solution using the appropriate values of $\alpha$ and $k$. The computed conductivities for all the above cases are in excellent agreement with the input values (Appendix A).

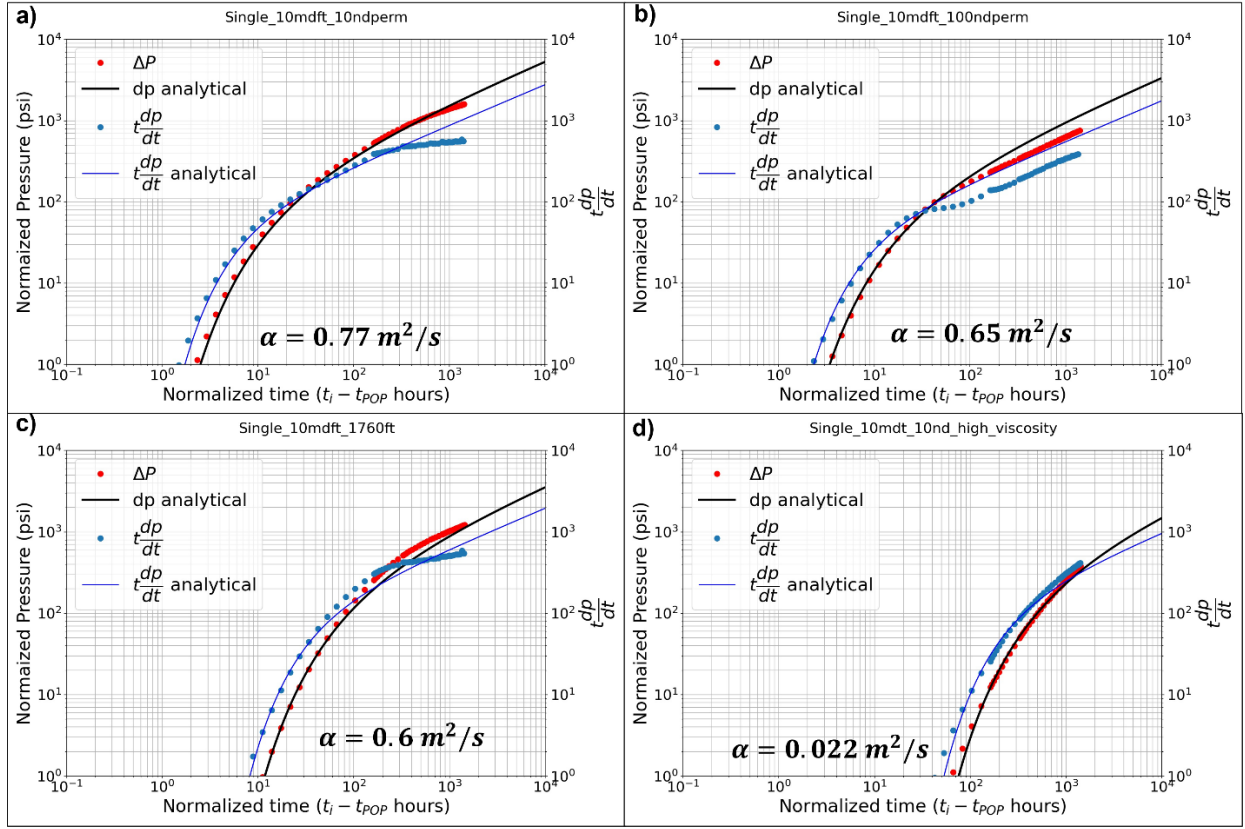

*Figure 7: Computing $\alpha$ for cases with varying permeability, well spacing and viscosity with a realistic pressure boundary condition in the Production Well. The cases are a) Reference: 10 mD-ft conductivity with 10 nD permeability and 880 ft well spacing; b) Reference conductivity with 100 nD permeability; c) 10 mD-ft conductivity with 1760 ft well spacing; d) 10 mD-ft cases with 30x higher viscosity. Note that the viscosity is accounted for while calculating conductivity from $\alpha$.*

### 4.1.4. Consistency between three interference tests

As discussed in Section 4.1.3, the baseline simulation setup includes three separate interference tests performed at different points in time. Because the fracture conductivity is constant over time during the simulations, we expect that if we interpret three interference tests, we should infer similar conductivities for each of them.

Figure 8 shows the plots of the three tests in the simulation with 10 mD-ft fracture conductivity and 10 nD permeability. All three tests show a similar value $\alpha$ for a good fit of the analytical solution to the observed $\Delta p$ and $t\frac{dp}{dt}$. Table 2 shows the estimated fracture conductivity for all three tests for three simulations with the input fracture conductivity values of 1, 10 and 100 mD-ft. In all three simulations, the estimated fracture conductivities are similar between the different tests and close to the input values.

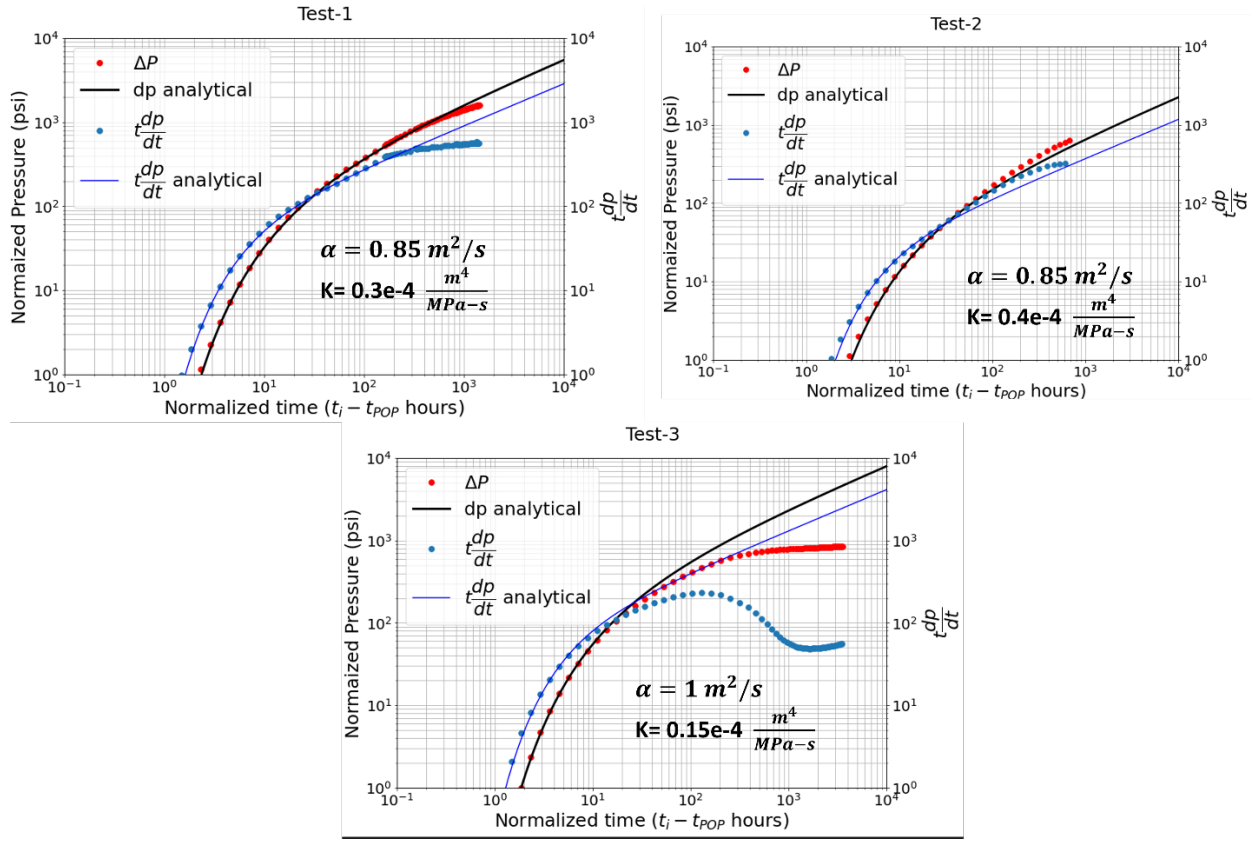

Figure 8. $\Delta p$, $t\frac{dp}{dt}$ from the analytical solution is plotted with the pressure observations from the three tests in the simulation model for the case with the 10 mD-ft fracture and 10 nD matrix permeability. The $\alpha$ values obtained for the three tests are consistent and provide a reasonable approximation of the input fracture conductivity.

Table 2. The calculated fracture conductivity ranges for the three tests are shown for cases with varying fracture conductivity.

| Case | Diffusivity ($\alpha$) ($\frac{m^2}{s}$) | | | Estimated fracture conductivity ranges (mD-ft) | | |
|---|---|---|---|---|---|---|
| | Test-1 | Test-2 | Test-3 | Test-1 | Test-2 | Test-3 |
| 10 mD-ft fracture with 10 nD permeability | 0.85 | 0.85 | 1 | 7.2-17.6 | 7.2-17.6 | 8.5-20.7 |
| 1 mD-ft fracture with 10 nD permeability | 0.06 | 0.06 | 0.07 | 0.5-1.2 | 0.5-1.2 | 0.6-1.4 |
| 100 mD-ft fracture with 10 nD permeability | 8 | 8 | 12 | 67.8-165.4 | 67.8-165.4 | 101.7-248.1 |

## 4.2. Estimating production impact from interference

While it is useful to estimate fracture conductivity, the ultimate objective of the interference tests is to understand the production impact from interference. The change in production rate at the Production Well when the Monitoring Well is shut-in (Figure 6) represents the production impact of the interference. As described in Section 1, this production impact can be quantified with the DPI metric.

Figure 9 shows the change in production rate for different simulations cases described in Section 4.1.3. As predicted by Equation 9, the relative production impact varies considerably, even for tests with the same conductivity. For example, Figure 9c shows that production impact increases with tighter well spacing. Figure 9d shows that production impact increases with lower viscosity.

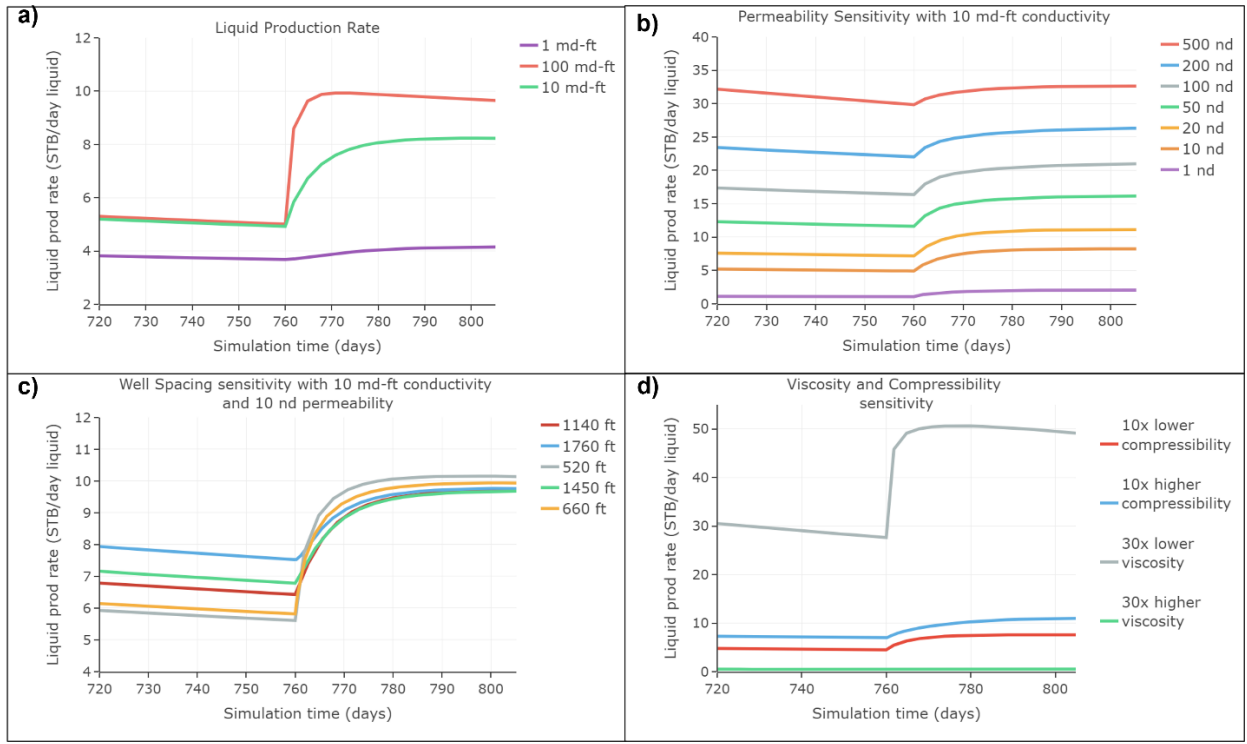

*Figure 9 a) production rate change with conductivity; b) production rate change with permeability; c) production rate change with well spacing; d) production rate change with viscosity and compressibility*

Following the procedure outlined in Section 3.4, we computed $L_D$ for all the simulation cases. We also computed the 24-hour CPG values for the same cases. Figure 10a shows the dimensionless interference length, and Figure 10b shows the 24-hour CPG plotted against the DPI for all of the simulations. Figure 10a shows that the simulations collapse onto a single curve. This demonstrates that the scaling in Equation 9 has successfully captured the relationship between the variables.

On the other hand, the plot of relative production impact versus CPG shows considerable scatter. In general, higher values of CPG result in greater values of DPI. However, we also find that the same value of CPG may be associated with substantially different values of DPI.

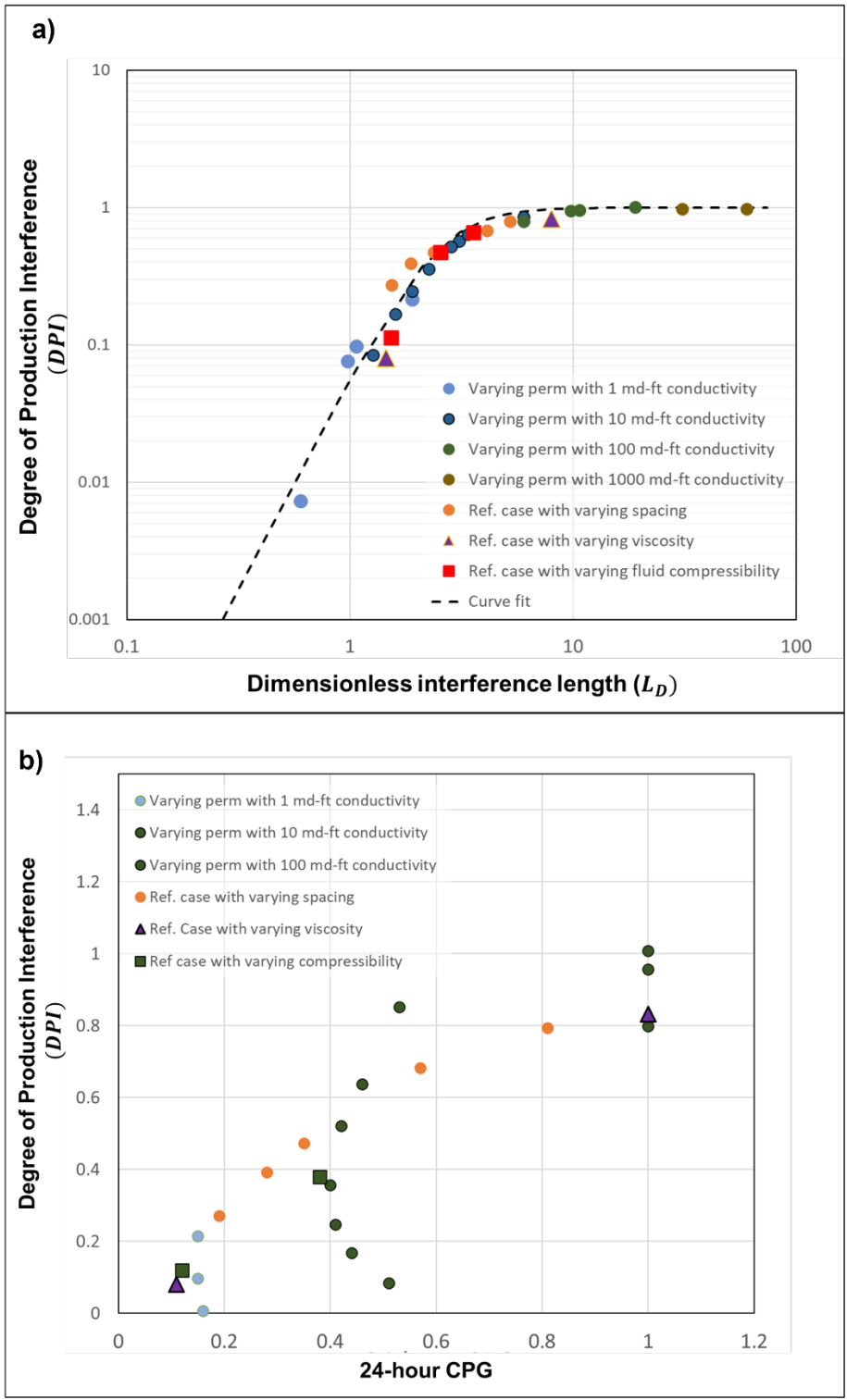

*Figure 10: a) DPI vs. $L_D$ is shown for the different cases. Each case represents a simulation with a particular combination of fracture conductivity, matrix permeability, well spacing, fluid viscosity and fluid compressibility. The reference case has a 10 mD-ft conductivity fracture, 10 nD matrix permeability, 880 ft well spacing, and water viscosity and compressibility. All the parameters are detailed in Table 2; b) DPI vs. 24-hour CPG for the same simulations. CPG values >1 have been truncated to 1 to be consistent with common practices in the industry.*

### 4.3. Cases with increasing complexity: multi-phase flow, multiple fractures, and heterogeneous conductivity

To further test the methodology, we ran additional cases to assess performance under more realistic conditions: multiple fractures, multi-phase flow, and a realistic distribution of proppant. The cases include:

1. Constant 10 mD-ft conductivity fracture with a dry gas reservoir
2. Constant conductivity fracture with a reservoir pressure that starts above the bubble point and eventually drops below the bubble point with gas coming out of solution. In this case, the pressure in the fracture is above the bubble point for Test-1, and then drops below the bubble point before the beginning of Test-2. The fracture conductivity is held constant at 10 mD-ft.
3. Realistic single stage simulation of a single stage with heterogenous distribution of proppant for an oil and gas reservoir.

Figure 11 shows the conductivity estimation for Case #1 with constant conductivity 10 md-ft fracture in a dry gas reservoir. The estimated conductivity is consistent with the input fracture conductivity in the simulation.

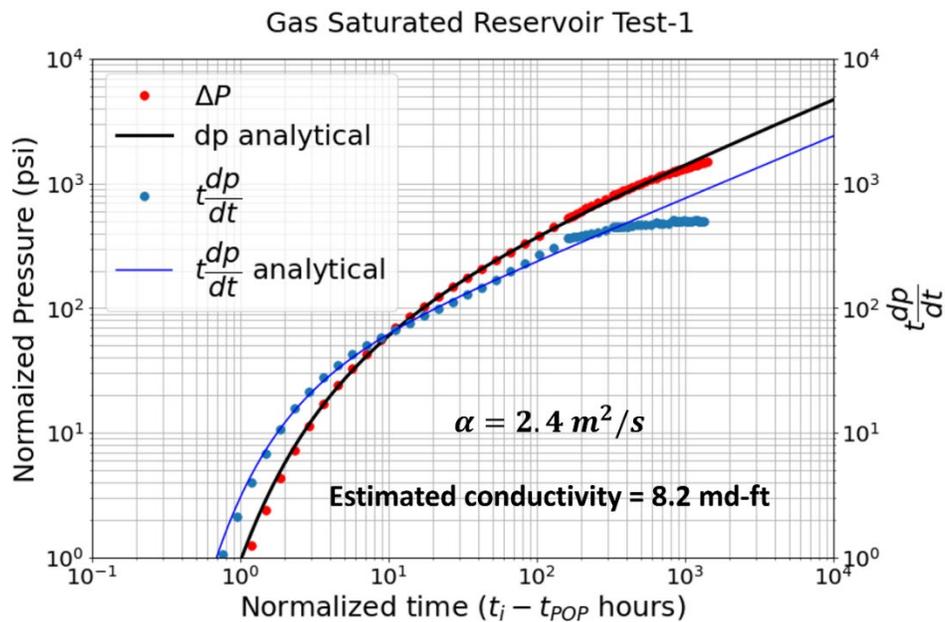

Figure 11. Conductivity estimation for Test-1 in Case#1.

Figure 12 shows the $dp$ and $t\frac{dp}{dt}$ plots for the three tests in Case #2. The analytical solution fits well, and the $\alpha$ values are consistent between the three tests. The multiphase mixture fluid viscosity and compressibility are calculated for each test by utilizing the method described in Appendix B. The computed conductivity for the three tests are: Test 1: 3-5 mD-ft, Test 2: 9-10 mD-ft, Test 3: 3-5 mD-ft. These results are reasonably consistent with the input fracture conductivity of 10 mD-ft. Figure 13 show the CPG plots for the three tests. The CPG values are reasonably consistent, 0.23, 0.19, and 0.26.

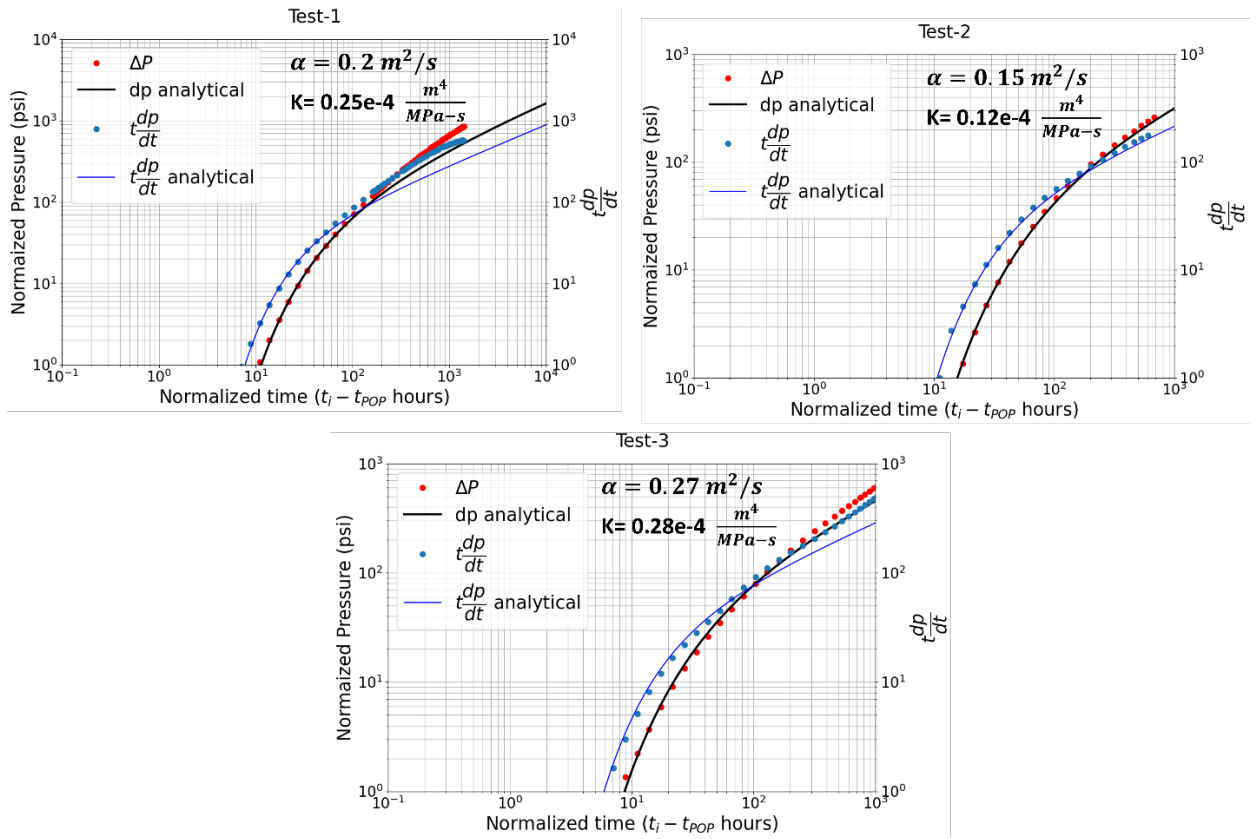

*Figure 12. Δp, $t\frac{dp}{dt}$ from the analytical solution is plotted with the pressure observations from the three tests in the simulation model for the case with the 10 mD-ft fracture with pressure above the bubble point during Test 1, which falls below the bubble point prior to Test 2 due to depletion caused the fluid withdrawal from the Production and Monitoring wells.*

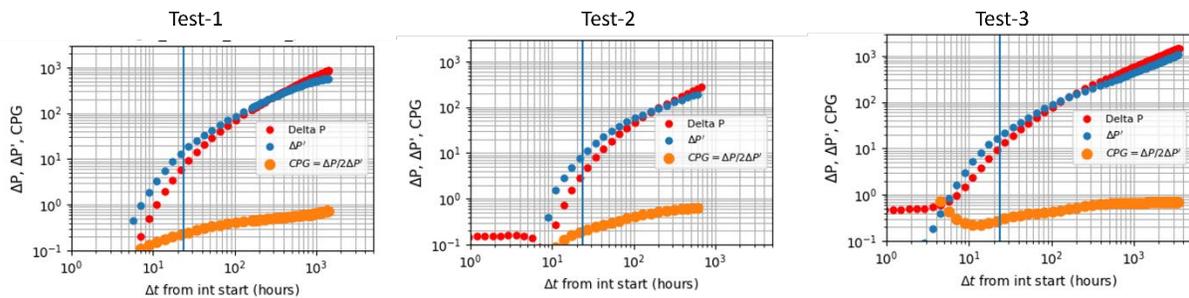

*Figure 13. CPG calculation for three tests in the simulation in which the pressure is above the bubble point pressure for Test-1 and falls below the bubble point pressure prior to the beginning of Test 2. The 24 hours CPG values for the three tests are 0.23, 0.19 and 0.26 respectively.*

The simulation for Case 3 uses a fully coupled 3-D hydraulic fracturing and reservoir simulation to represent a typical single hydraulic fracturing stage from a real case study with 10 perforation cluster spaced 20 ft apart. The simulator uses a fully coupled 3-D approach to model fracture propagation, leak-off, stress shadow and proppant distribution within the fracture (McClure et al., 2022c).

Figure 14a shows the fluid and proppant injection schedule input into the simulations for both the Production and Monitoring wells. Figure 14b shows the proppant distribution in the connected fracture between the two wells. While the distribution of the proppant is heterogenous, the two wells are fully connected by multiple fractures with overlapping propped areas. Additional details of the realistic single stage simulation are given in Appendix C.

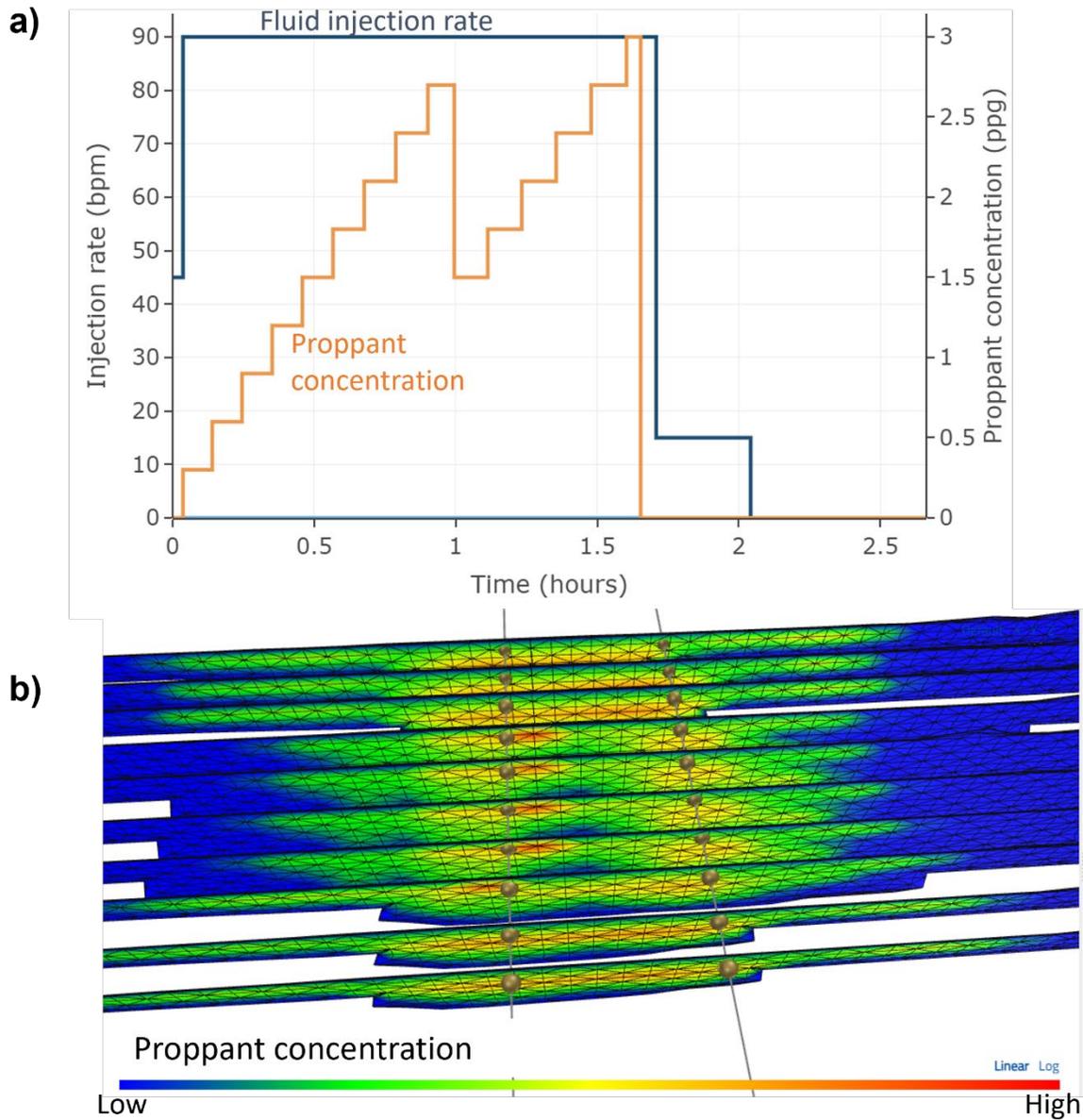

*Figure 14. a) Pump schedule used for the realistic hydraulic fracture stage simulation; b) Proppant distribution in the resulting connected fractures.*

Figure 15 shows that the results from the three cases with increasing complexity continue to collapse onto the same curve as seen from the results in Section 4.2. These results demonstrate that the method is robust, even when applied to fully realistic cases.

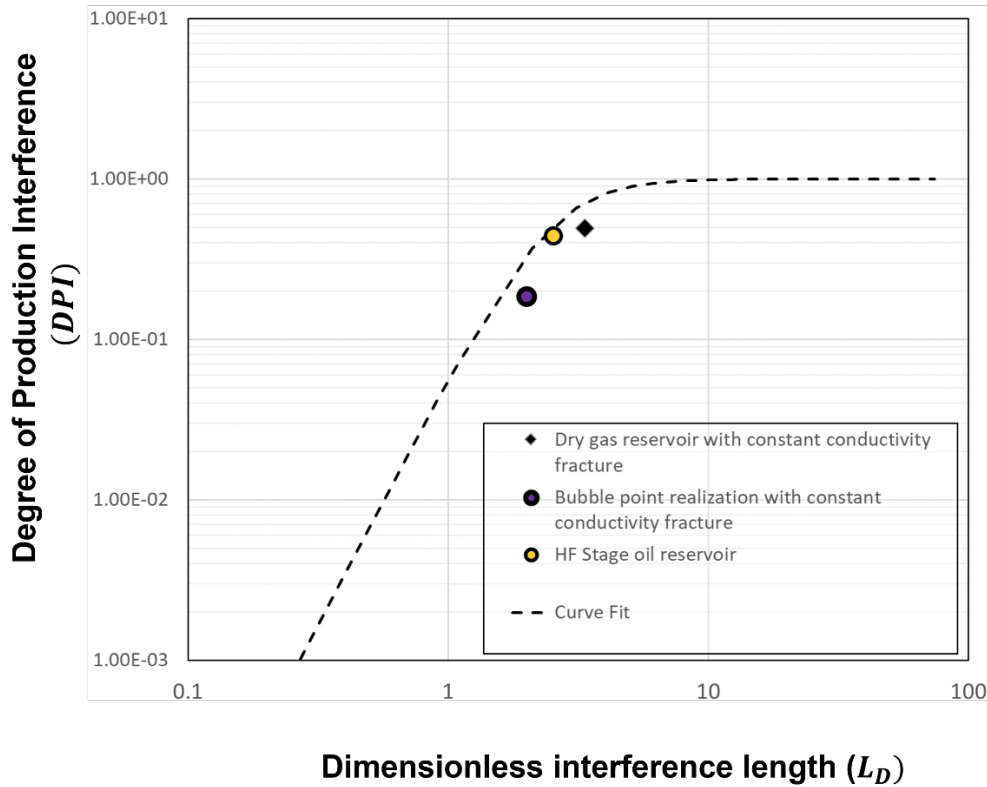

*Figure 15. DPI vs. $L_D$ with more complex cases included.*

The topic of heterogenous fracture conductivity deserves investigation in future work. If conductivity between the wells is controlled by one or a small number of highly conductive fractures, then this has the potential to bias the results.

## 5. Demonstrating the DQI method using real-life field examples

To further test DQI, we applied the method to two field tests with multiple interference tests, varying fluid properties, multi-wells, multi-benches, and varying completion designs. The setup of the two tests along with the results will be discussed in this section.

### 5.1. Field Example from the Anadarko Basin

We applied the DQI procedure to interpret a series of interference tests conducted in a pad with four hydraulically fractured wells in the Meramec formation of the Anadarko basin. Miller et al. (2019) provides a detailed overview of the Meramec geology in Oklahoma. Figure 16a shows the gun barrel view of the well locations and Figure 16b show shows the sequence of operations. A preexisting parent well labelled as Well P had been producing for several months prior to the interference testing campaign. The following interference tests were analyzed:

1. Well 1 POP: BHP changes observed in Wells 2,3 and 4.
2. Well 2 POP: BHP changes observed in Wells 3 & 4.
3. Well 3 POP: BHP changes observed in Well 4.

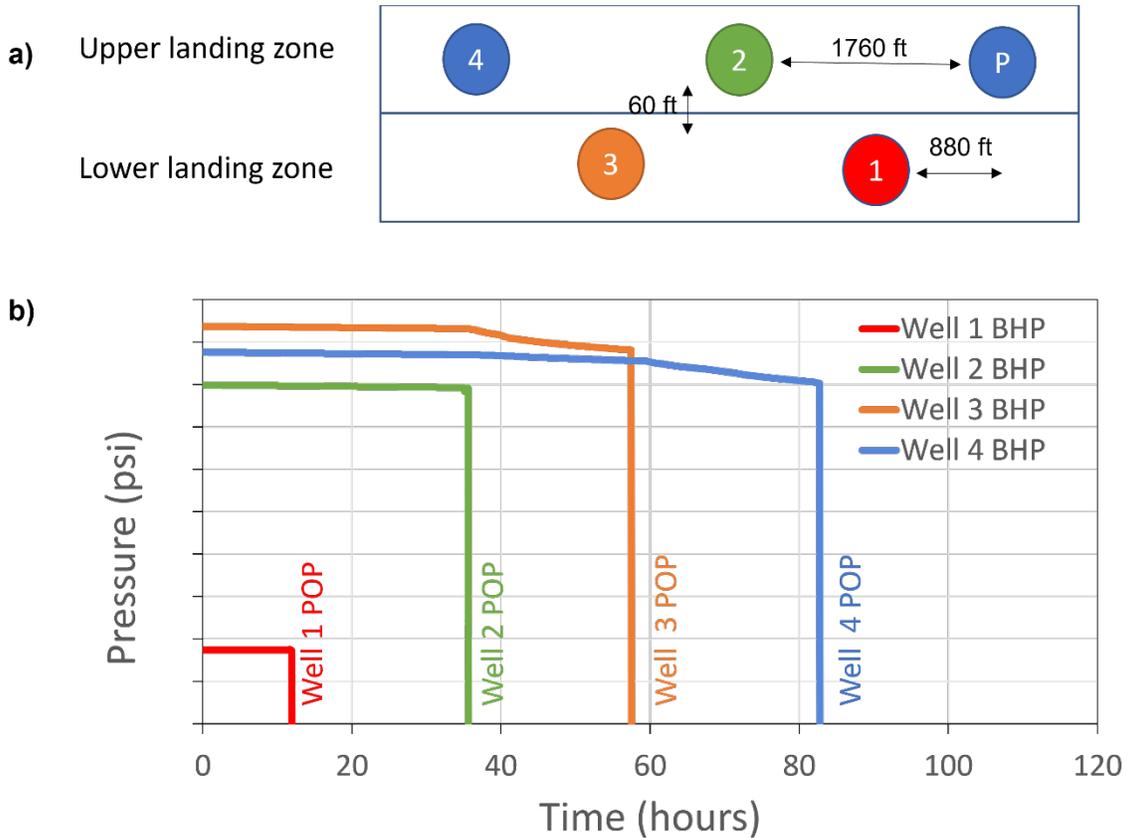

*Figure 16 a) Gun barrel view of the well locations; b) Sequence of well POP and the BHP changes in the Monitoring Wells.*

Following the procedure outlined in Section 3, the DQI analysis was performed. Figure 13 shows the extrapolation of the prior pressure trends to calculate $p_{ref}$. Figure 14 shows the curve fit of the analytical linear flow solution on the Bourdet plots. Table 4 shows the inferred values of hydraulic diffusivity, conductivity, and DPI. The CPG estimate is also shown for comparison.

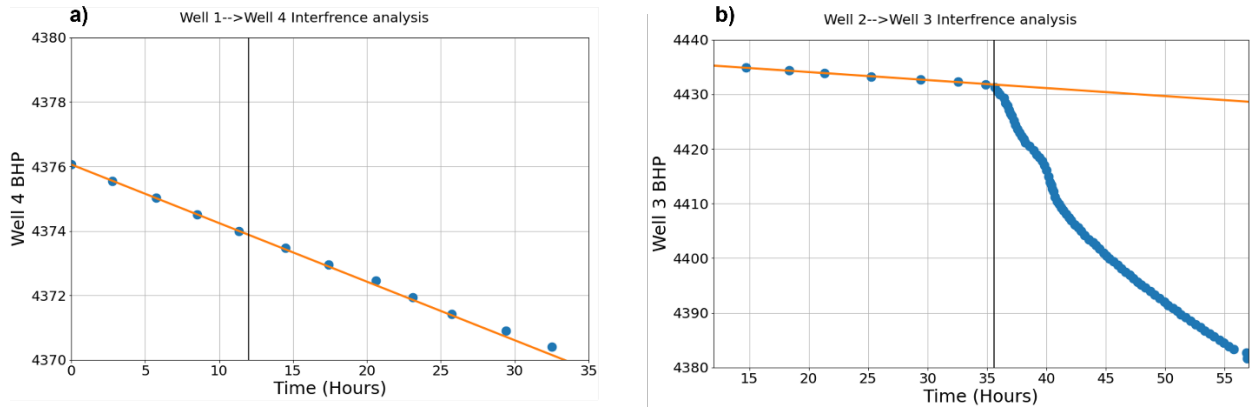

*Figure 17 a) Interference tests between Well 1 and Well 4 showing negligible interference; b) Interference test between Well 2 and Well 3 showing significant interference.*

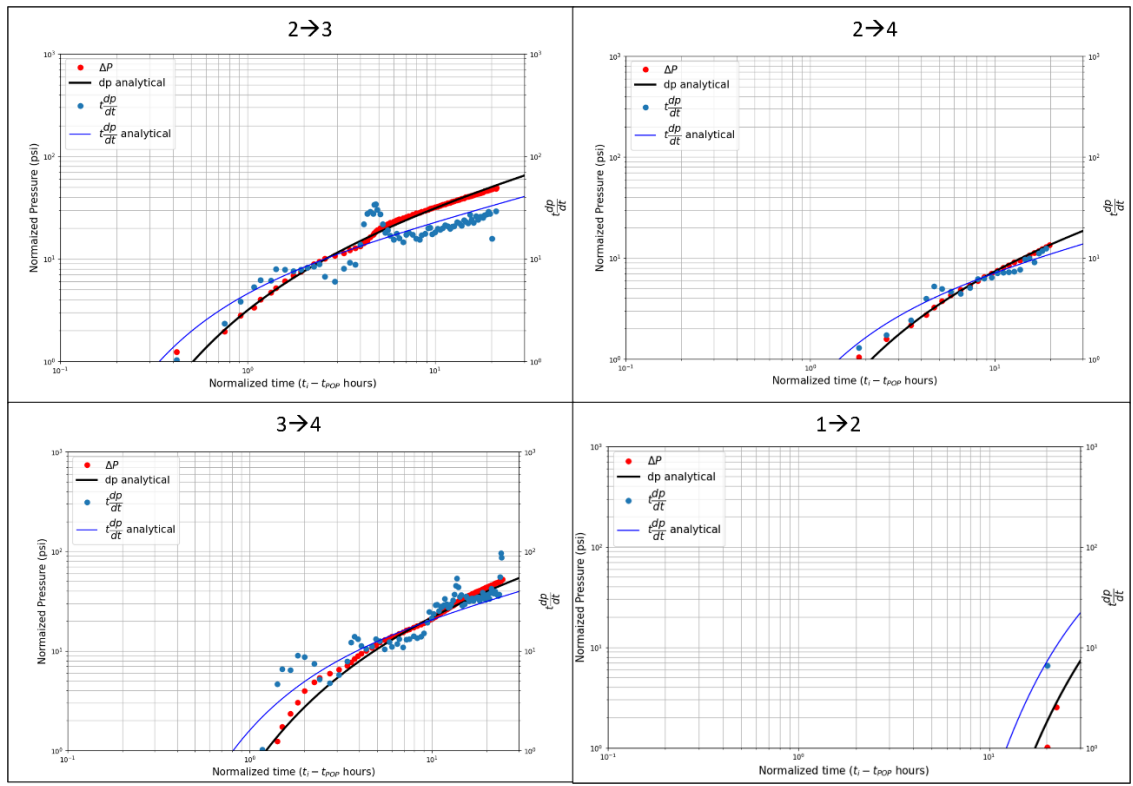

*Figure 18 Fitting on the analytical solution to the 4 interference tests.*

*Table 3. Summary of interfrence tests in the Meramec formation*

| Test | Fracture conductivity | Dimensionless interference length ($L_D$) | CPG | 20-day DPI predicted from the $L_D$ estimate |
|---|---|---|---|---|
| 1→2 | 0.4-0.6 mD-ft | 0.5-0.6 | 0.45-0.5 | >1% |
| 1→3 | No interference | | | 0% |
| 1→4 | No interference | | | 0% |
| 2→3 | 60-90 mD-ft | 6.0-7.3 | 0.8-1 | ~70% |
| 2→4 | 10-20 mD-ft | 1.2-1.7 | 0.5-0.7 | 20-30% |
| 3→4 | 10-20 mD-ft | 2.4-3.5 | 0.4-0.9 | 40-50% |

It is very important to note that the computed DPIs are indicative of the interference at the time of the tests. Fracture conductivity is expected to significantly reduce during drawdown and long-term production (Li et al., 2020), and this will reduce the DPI. Thus, to track changes over time, it would be useful to perform multiple interference tests at different points in time. Interference tests performed after at least months of production will yield a conductivity and DPI estimate that is representative of the actual *long term* production interference between the wells.

The CPG and DPI estimates are qualitatively consistent, with the exception of the interference from Well 1 to Well 2 (lower right plot of Figure 18). In this test, Figure 18 shows that this test had a delayed and low-magnitude pressure response. Consequently, the estimates for conductivity and DPI are low, suggesting that the connection is relatively weak. However, the CPG metric for this interference response is similar to the CPG metric for Wells 2 and 3 to Well 4, even though those responses occurred much more rapidly and with greater magnitude. The interpretation from the DPI metric appears more physically plausible than the interpretation from the CPG estimate.

The methods can yield qualitative different results because the CPG method is based on the power law scaling of the *shape of the curve*, rather than the timing and magnitude of the curve.

## 5.2. Field Example from the Delaware Basin

In the second case study, we applied the DQI method to a series of interference tests conducted with 6 wells in the Delaware basin in three separate landing zones. Four of the 6 wells had bottom hole pressure gauges. Figure 19a shows the gun barrel view of the well locations, and Figure 19b shows the order of the wells put on production. A total of 13 interference tests were interpreted. In contrast to the Meramec dataset, the production had a very high water cut in the early production sequence and the analysis could be simplified by assuming water to the single interfering phase.

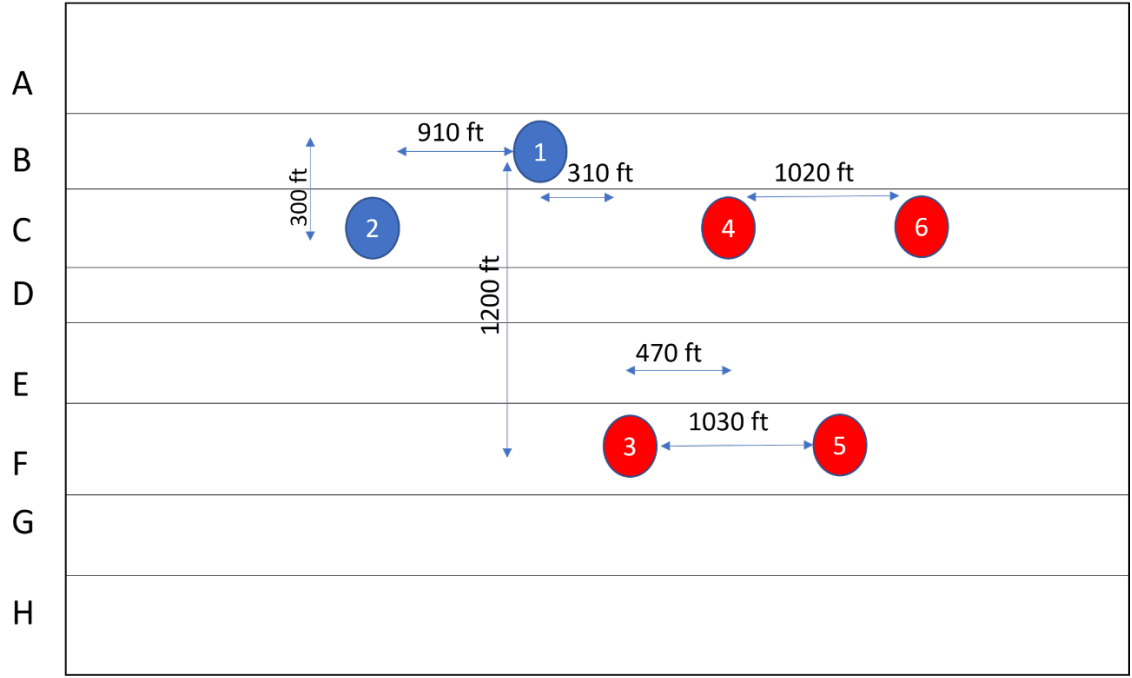

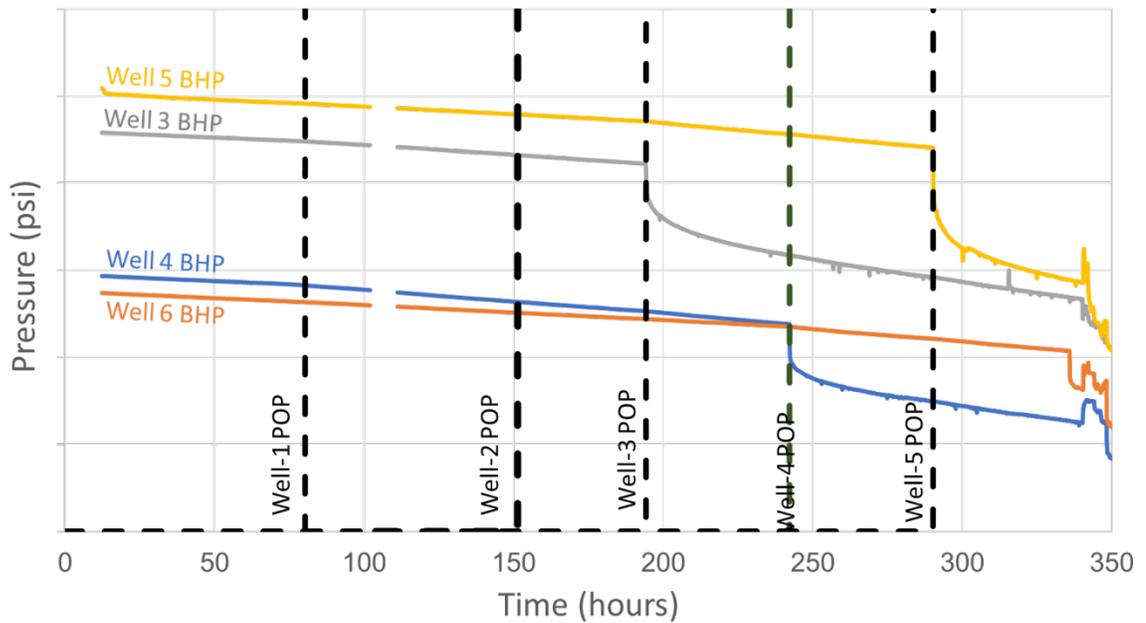

Figure 19. a) Gun barrel view of the well locations and the relative distances. Wells 3,4,5 and 6 in red were monitored with pressure gauges; b) Sequence of POP of the 6 wells and the BHP in 4 well with the gauges

We analyzed this case study using the procedure outlined in Section 3. 9 out of 14 interference tests showed no interference. For example, the Well 2 POP showed no interference signal in any of the 4 monitoring wells due to the large distance. Figure 20a and b show examples of two interference tests in which noticeable interference was detected and the data was fit with the analytical solution from Equation 6. The solution was able to fit the data from all the interference tests reasonably well. Figure 20c summarizes the fracture conductivity estimates for the pairs of POP and monitoring wells in which significant interference was detected. Wells in the same geological layer have a clearly stronger propped connection than wells across different landing zones. Table 4 summarized the estimated $L_D$ and DPI values for these analyzed interference tests.

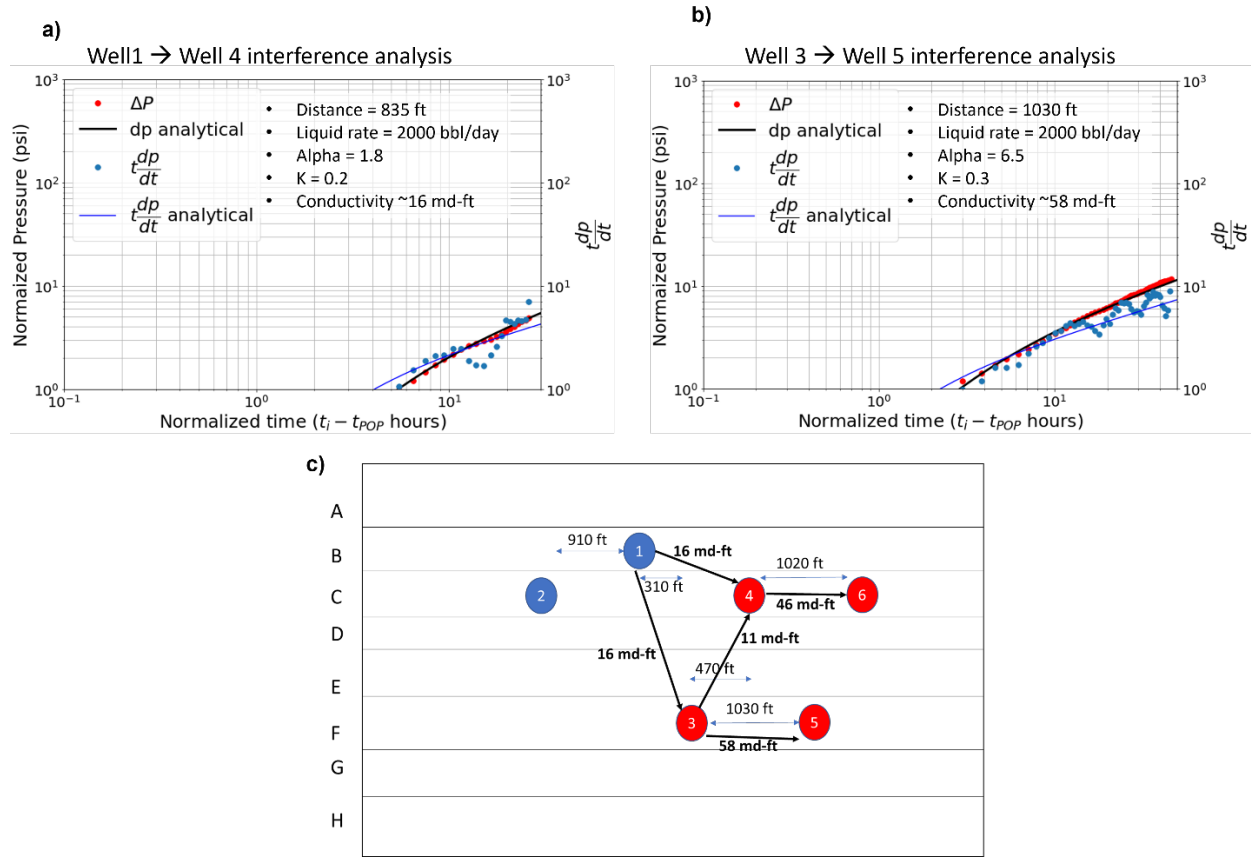

Figure 20 a), b) Analytical solution fit examples from two interference tests; c) Summary of the fracture conductivity values estimated from all the tests.

Table 4 Summary of the Delaware basin interference tests which showed interference signals

| Test | Fracture conductivity (mD-ft) | Dimensionless interference length ($L_D$) | Estimated 20-day DPI |
|---|---|---|---|
| 1→3 | 15-35 | 2.0-3.0 | 30-40% |
| 1→4 | 15-35 | 2.9-4.5 | 60-70% |
| 3→4 | 10-25 | 2.0-3.0 | 30-40% |
| 4→6 | 45-105 | 4.1-6.3 | 80-90% |
| 3→5 | 55-135 | 4.5-7.1 | 80-90% |

Again, we should caution that these values of DPI and conductivity are derived based on the fracture conductivity when the wells are initially put on production. The conductivity is likely to reduce by 10-100x during long-term drawdown, which will result in significantly lower values of DPI. In Table 5, we show the impact of hypothetically reducing the fracture conductivity by 10x and 40x on the DPI to represent the loss in conductivity during long-term drawdown.

Table 5. Estimated DPI with loss with a hypothetical 10x and 40x loss in fracture conductivity due to long-term drawdown.

| Test | Estimated DPI with Fracture Conductivity divided by 10 | Estimated DPI with Fracture Conductivity divided by 40 |
|---|---|---|
| 1→3 | 1-5% | <1% |
| 1→4 | 5-15% | <1-2% |
| 3→4 | 1-5% | <1% |
| 4→6 | 12-35% | 1-5% |
| 3→5 | 15-40% | 2-7% |

## 6. Discussion – interpretation of CPG values

The simulation results provide intuition that can be helpful for understanding the physical significance of the CPG metric.

The CPG value is calculated from the power-law scaling of pressure with time. The general power law equation can be written as:

$$\Delta p = At^a \quad (12)$$

where $a$ is the power-law exponent. From the definition of CPG in Equation 3, the CPG value can be related to the power law exponent as:

$$\text{CPG} = \frac{1}{2a} \quad (13)$$

Linear flow (exponent of 0.5) has a CPG value of 1.0. Bilinear flow (exponent of 0.25) yields a CPG value of 2.0. CPG values less than one suggest that pressure is scaling with an exponent greater than 0.5; an exponent of 1.0 (linear with time) implies a CPG of 0.5.

It may seem somewhat paradoxical that lower values of CPG imply weaker connection between wells, because lower CPG implies a larger power-law decay with time (i.e., larger power law exponent in Equation 12). The simulation results suggest that this may occur because of the timing of the CPG measurement with respect to the onset of pressure interference and the transition into the long-term transient.

Figure 21 shows CPG, $\Delta p$, and the $t\frac{dp}{dt}$ curves for the very simple simulated interference tests from Section 4.1. The CPG observations from these tests are summarized in Table 6. Prior to the onset of interference, $\Delta p$ is zero, and so the curve is not visible. Panels (a) and (b) correspond to simulations with relatively low values of conductivity, and hours pass until a measurable pressure change is detected. Once the onset of pressure change occurs, the slope of the $t\frac{dp}{dt}$ curve is temporarily very steep, as it rises from zero and eventually settles into its long-term trend. This steep-to-shallow $t\frac{dp}{dt}$ trend on the log-log plot is also apparent in the analytical solution for 'linear flow observed at an offset' (Figure 2).

The slope of the $t\frac{dp}{dt}$ curve is equal to the power law exponent in Equation 3 which starts very steep and flattens. Thus, because CPG is equal to $1/(2\alpha)$, the high-to-decreasing trend in the apparent power law exponent corresponds to a low-to-high trend for CPG. Correspondingly, in Figure 21, CPG rises from zero at the onset of interference, and gradually approaches its long-term value. In this particular simulation, the long-term flow-regime is matrix linear flow, and so the CPG gradually approaches 1.0.

CPG is typically measured at around 24 to 48 hours after the start of the test. Tests with stronger interference are 'further along' in the transition from low-to-high, and so tend to yield higher values of CPG.

This discussion is probably an oversimplification. In reality, CPG may be affected by a variety of other factors, such as the rate of pressure decline at the production well, matrix permeability, the potential for anomalous diffusion (Raghavan & Chen, 2013; Ren & Guo, 2015). Furthermore, CPG may be affected by mechanisms which impact the power law assumption such as fracture skin (Jha & Lee, 2022).

Overall, our results suggest that when comparing wells under similar reservoir and test conditions, the CPG metric is usually reasonably successful at qualitatively comparing relative connectivity. However, because the metric is affected by a variety of additional factors (Figure 10), CPG values from different formations or under different test conditions may not be comparable.

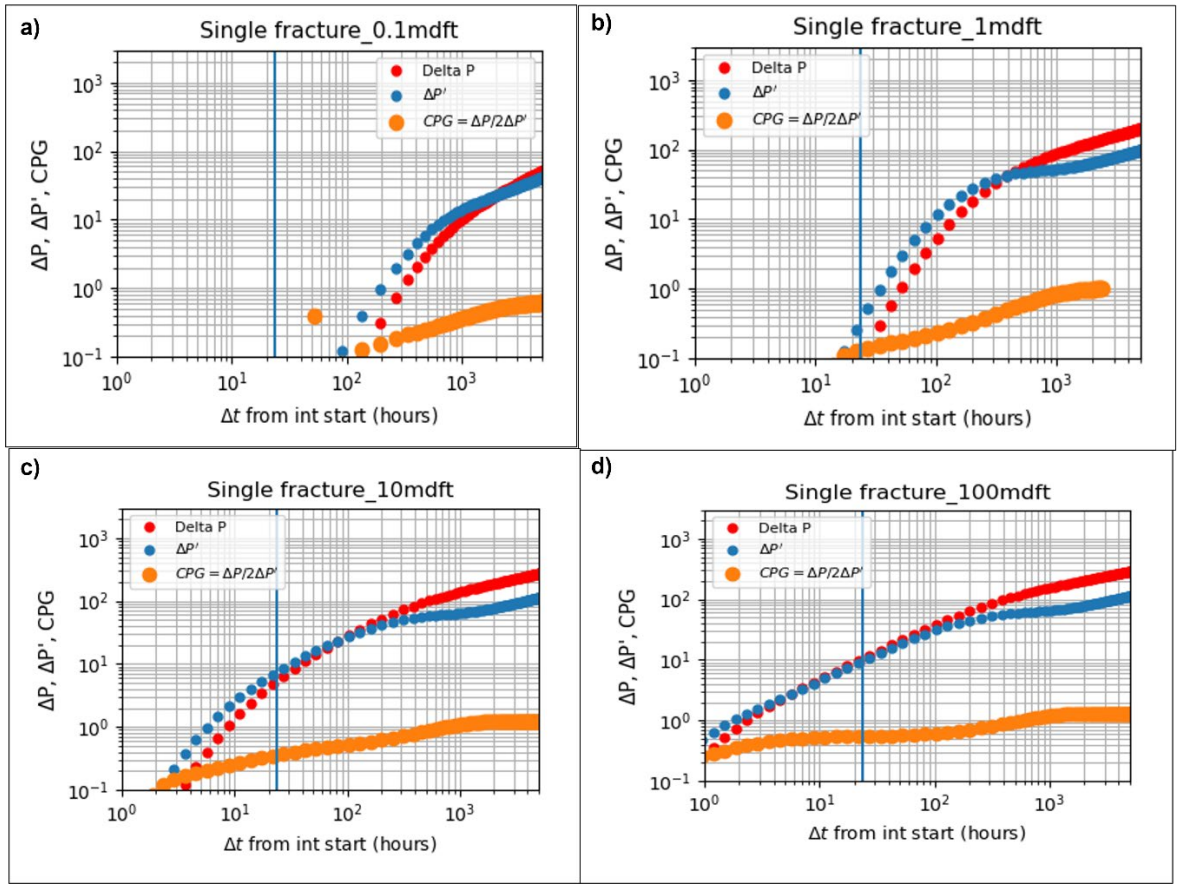

Figure 21. CPG versus time for the constant rate simulations.

Table 6. CPG and $t\frac{dp}{dt}$ observations for the extended interference tests

| Fracture Conductivity (mD-ft) | CPG values | | $t\frac{dp}{dt}$ observations | | Comments |
|---|---|---|---|---|---|
| | 24-hour | Long term | 24-hour | Long term | |
| 0.1 | NA | 0.7 and rising | NA | 0.5-1 | ½ slope not reached in the test |
| 1 | 0.1 | ~1 | ~1 | 0.5 | Final half slope reached in ~1.3months |
| 10 | 0.35 | 1.1-1.2 | ~1 | 0.5 | Final half slope reached in ~1.2 months |
| 100 | 0.55 | 1.2-1.3 | ~1 | 0.5 | Final half slope reached in ~1-1.2 months |

# 7. Conclusions

We present the newly developed DQI method for analyzing interference tests between wells in shale. The DQI provides subsurface engineers with a practical procedure to maximize the value of downhole pressure gauges and better optimize well spacing and completions design on future development programs.

The procedure starts by estimating the hydraulic diffusivity from the initial onset of interference at the observation well. Then, the diffusivity is used to estimate hydraulic conductivity, which is used to estimate a dimensionless drainage length parameter, $L_D$. We introduce a metric based on production rate changes to estimate the degree of production interference, the DPI. Based on simulations under a variety of conditions, we empirically develop a relationship between DPI and $L_D$.

The advantages of DQI are that:

a. It is robust to differences in drawdown schedule and/or nonlinearities in the fracture flow because it relies on the initial onset of pressure interference at the observation well.
b. It relates the observations to fracture conductivity, a physical parameter that would otherwise be difficult to measure.
c. It provides a framework for estimating how production will be impacted by interference.
d. It takes into account the effects of varying the fluid and rock properties.

DQI is the first attempt in the industry at combining concepts from pressure transient analysis, numerical simulation, and field data to characterize pressure responses of interference tests by relating them to the physics of fluid flow and porous media.

In future work, we hope to: (a) program DQI in Python to automate certain aspects of the workflow, (b) apply the procedure to a much larger number of field datasets, and evaluate its performance, (c) use simulations to evaluate whether further additional information can be extracted from interference tests, and (d) consider integrating with interference observations from fiber.

In addition to interference test interpretation, the relationship between $L_D$ and DPI could also be used directly in well spacing optimization.

## Acknowledgements

The authors would like to thank Devon Energy for supporting this project. We are also thankful to all of the execution engineers and field personnel who led the deployment of downhole pressure gauges used in this project.

# Appendix A: Table of raw results

Table 7: Fracture conductivity calculations for all simulation cases with the prescribed pressure decline shown in Figure 6, which is consistent with typical shale well pressure drawdowns.

| Parameter | Simulation input parameters | | | | | Estimated from data | Assumed reasonable values | | | Estimated from diffusivity ($\alpha$) | |
|---|---|---|---|---|---|---|---|---|---|---|---|
| | Conductivity | Permeability | Viscosity | Compressibility | Well spacing | Diffusivity ($\alpha$) | dW/dp (m/MPa) | | Aperture | Fracture conductivity | |
| | | | | | | | low | high | W | low | high |
| | md-ft | nd | cp | psi^-1 | ft | m2/s | m/MPa | m/MPa | m | md-ft | md-ft |
| Conductivity | 1 | 15 | 0.31 | 0.000003 | 880 | 0.07 | 8.00E-06 | 2.00E-05 | 7.60E-04 | 0.6 | 1.4 |
| | 10 | 15 | 0.31 | 0.000003 | 880 | 0.75 | 8.00E-06 | 2.00E-05 | 7.60E-04 | 6.4 | 15.5 |
| | 100 | 15 | 0.31 | 0.000003 | 880 | 7.2 | 8.00E-06 | 2.00E-05 | 7.60E-04 | 61.0 | 148.9 |
| | 1000 | 15 | 0.31 | 0.000003 | 880 | 70 | 8.00E-06 | 2.00E-05 | 7.60E-04 | 593.1 | 1447.4 |
| Permeability | 10 | 1 | 0.31 | 0.000003 | 880 | 0.9 | 8.00E-06 | 2.00E-05 | 7.60E-04 | 7.6 | 18.6 |
| | | 10 | 0.31 | 0.000003 | 880 | 0.85 | 8.00E-06 | 2.00E-05 | 7.60E-04 | 7.2 | 17.6 |
| | | 20 | 0.31 | 0.000003 | 880 | 0.77 | 8.00E-06 | 2.00E-05 | 7.60E-04 | 6.5 | 15.9 |
| | | 50 | 0.31 | 0.000003 | 880 | 0.68 | 8.00E-06 | 2.00E-05 | 7.60E-04 | 5.8 | 14.1 |
| | | 100 | 0.31 | 0.000003 | 880 | 0.65 | 8.00E-06 | 2.00E-05 | 7.60E-04 | 5.5 | 13.4 |
| | | 200 | 0.31 | 0.000003 | 880 | 0.62 | 8.00E-06 | 2.00E-05 | 7.60E-04 | 5.3 | 12.8 |
| | | 500 | 0.31 | 0.000003 | 880 | 0.6 | 8.00E-06 | 2.00E-05 | 7.60E-04 | 5.1 | 12.4 |
| | 1 | 1 | 0.31 | 0.000003 | 880 | 0.07 | 8.00E-06 | 2.00E-05 | 7.60E-04 | 0.6 | 1.4 |
| | | 10 | 0.31 | 0.000003 | 880 | 0.06 | 8.00E-06 | 2.00E-05 | 7.60E-04 | 0.5 | 1.2 |
| | | 100 | 0.31 | 0.000003 | 880 | 0.06 | 8.00E-06 | 2.00E-05 | 7.60E-04 | 0.5 | 1.2 |
| | 100 | 1 | 0.31 | 0.000003 | 880 | 7 | 8.00E-06 | 2.00E-05 | 7.60E-04 | 59.3 | 144.7 |
| | | 10 | 0.31 | 0.000003 | 880 | 8 | 8.00E-06 | 2.00E-05 | 7.60E-04 | 60.2 | 146.8 |
| | | 100 | 0.31 | 0.000003 | 880 | 7.1 | 8.00E-06 | 2.00E-05 | 7.60E-04 | 67.8 | 165.4 |
| Viscosity | 10 | 10 | 0.010333 | 0.000003 | 880 | 22 | 8.00E-06 | 2.00E-05 | 7.60E-04 | 6.2 | 15.2 |
| | | 10 | 9.3 | 0.000003 | 880 | 0.022 | 8.00E-06 | 2.00E-05 | 7.60E-04 | 5.6 | 13.6 |
| Compressibility | 10 | 10 | 0.31 | 0.00003 | 880 | 0.5 | 8.00E-06 | 2.00E-05 | 7.60E-04 | 5.7 | 11.9 |
| | | 10 | 0.31 | 0.0003 | 880 | 0.1 | 8.00E-06 | 2.00E-05 | 7.60E-04 | 4.2 | 5.4 |
| Well spacing | 10 | 15 | 0.31 | 0.000003 | 520 | 1.35 | 8.00E-06 | 2.00E-05 | 7.60E-04 | 11.4 | 27.9 |
| | | 15 | 0.31 | 0.000003 | 600 | 0.9 | 8.00E-06 | 2.00E-05 | 7.60E-04 | 7.6 | 18.6 |
| | | 15 | 0.31 | 0.000003 | 1140 | 0.8 | 8.00E-06 | 2.00E-05 | 7.60E-04 | 6.8 | 16.5 |
| | | 15 | 0.31 | 0.000003 | 1450 | 0.8 | 8.00E-06 | 2.00E-05 | 7.60E-04 | 6.8 | 16.5 |
| | | 15 | 0.31 | 0.000003 | 1760 | 0.6 | 8.00E-06 | 2.00E-05 | 7.60E-04 | 5.1 | 12.4 |

# Appendix B: Calculation of fluid properties with multi-phase fluid flow

To estimate $L_D$, we need to estimate the fluid compressibility, viscosity, and total mobility in the reservoir. These values cannot be known with high precision. However, we can use reasonable simplifying assumptions to derive a first-order approximation.

First, we seek to estimate the fluid saturations in the formation. If using the black oil model, we can convert surface volumes to reservoir volumes:

$$RB_o = STB_o B_o, \tag{B1}$$

$$RB_g = (scf_g - R_s STB_o) B_g, \tag{B2}$$

$$RB_w = STB_w B_w, \tag{B3}$$

If using the modified black oil model, then Equations B1 and B2 must be replaced with:

$$STB_o = \frac{RB_o}{B_o} + \frac{RB_g}{B_g} R_v, \tag{B4}$$

$$scf_g = \frac{RB_g}{B_g} + \frac{RB_o}{B_o} R_s. \tag{B5}$$

Then, Equations B4 and B5 must be solved jointly for $RB_g$ and $RB_o$.

The values from the black oil table, $B_o$, $B_g$, $R_s$, and $R_v$, can be evaluated at the production well's bottomhole pressure. If the BHP is greater than the saturation pressure, then we can safely assume that there is only one hydrocarbon phase present in the reservoir. $B_w$ is nearly always close to 1.0.

$RB$ stands for 'reservoir barrels,' or more generally, the volume of the phase in the reservoir. These $RB$ values should be interpreted as the flowing volumes in the reservoir, not the actual 'in-place' volumes.

The total mobility is defined as:

$$M_t = \frac{k_{rw}}{\mu_w} + \frac{k_{ro}}{\mu_o} + \frac{k_{rg}}{\mu_g}. \tag{B6}$$

We can write that:

$$\frac{1}{M_t} \frac{k_{rw}}{\mu_w} = \frac{RB_w}{RB_w + RB_o + RB_g} \tag{B7}$$

$$\frac{1}{M_t} \frac{k_{ro}}{\mu_o} = \frac{RB_o}{RB_w + RB_o + RB_g} \tag{B8}$$

$$\frac{1}{M_t} \frac{k_{rg}}{\mu_g} = \frac{RB_g}{RB_w + RB_o + RB_g} \tag{B9}$$

Equations B6-B9 define a system of equations that can be solved for the relative permeability of the three phases. The viscosity values can be read from the black oil table. In shale, typical values are $\mu_w = 0.3\ cp$, $\mu_o = 0.3\ cp$, and $\mu_g = 0.03\ cp$.

If we have estimates for the relative permeability curves, we can solve them to infer phase saturations from the relative permeabilities. Alternatively, you may use a simple assumption, such as:

$$k_{rp} = S_p \tag{B10}$$

Or:

$$k_{rp} = (S_p - 0.2)^2 \tag{B11}$$

With knowledge of the saturations, the total compressibility can be calculated as:

$$c_t = c_\phi + S_w c_w + S_o c_o + S_g c_g \tag{B12}$$

The value of $c_\phi$ can be assumed to be 1e-5 psi$^{-1}$. The value of $c_w$ can be assumed to be 3e-6 psi$^{-1}$. The oil and gas phase compressibilities can be estimated from the black oil table values:

$$c_g = -\frac{1}{B_g}\frac{dB_g}{dp} \tag{B13}$$

$$c_o(above\ bubble\ point) = -\frac{1}{B_o}\frac{dB_o}{dp} \tag{B14}$$

$$c_o(below\ bubble\ point) = -\frac{1}{B_o}\left[\left(\frac{dB_o}{dp}\right) - B_g\left(\frac{dR_s}{dp}\right)\right] \tag{B15}$$

Typical values are: $c_g$ ~ 1/p, $c_o$ ($p > p_{sat}$) ~ 1e-5 psi$^{-1}$, and $c_o$ ($p < p_{sat}$) ~ 2e-4 psi$^{-1}$.

# Appendix C: Details of the realistic stage simulation

The simulation of a realistic stage in Section 4.3 was performed utilizing a fully coupled 3D hydraulic fracture, fluid flow and geomechanics simulator. The pump design shown in Figure 14a was used for both the Production and Monitoring wells. The formation fluid and reservoir properties are consistent with a history matched dataset from the Meramec formation. The injected fluid was set to be a high viscosity friction reducer (HVFR) with fluid viscosity varying between 10-35 cP as a function of strain rate. The interfering fluid during Test-1 is a mixture of the HVFR and formation fluid as the fracture is still filled with the injected fluid. Test-1 involves a fluid mixture with a heterogenous distribution of viscosity between the two end member extremes. Test-2 and Test-3 only involve formation fluids with the viscosity and compressibility calculated using the methodology outlined in Appendix B. The fracture aperture and consequently conductivity is a function of the effective normal stress and varies between the tests. Figure 22 shows the 3D plots of the fractures with the fluid viscosity, fracture conductivity and aperture.

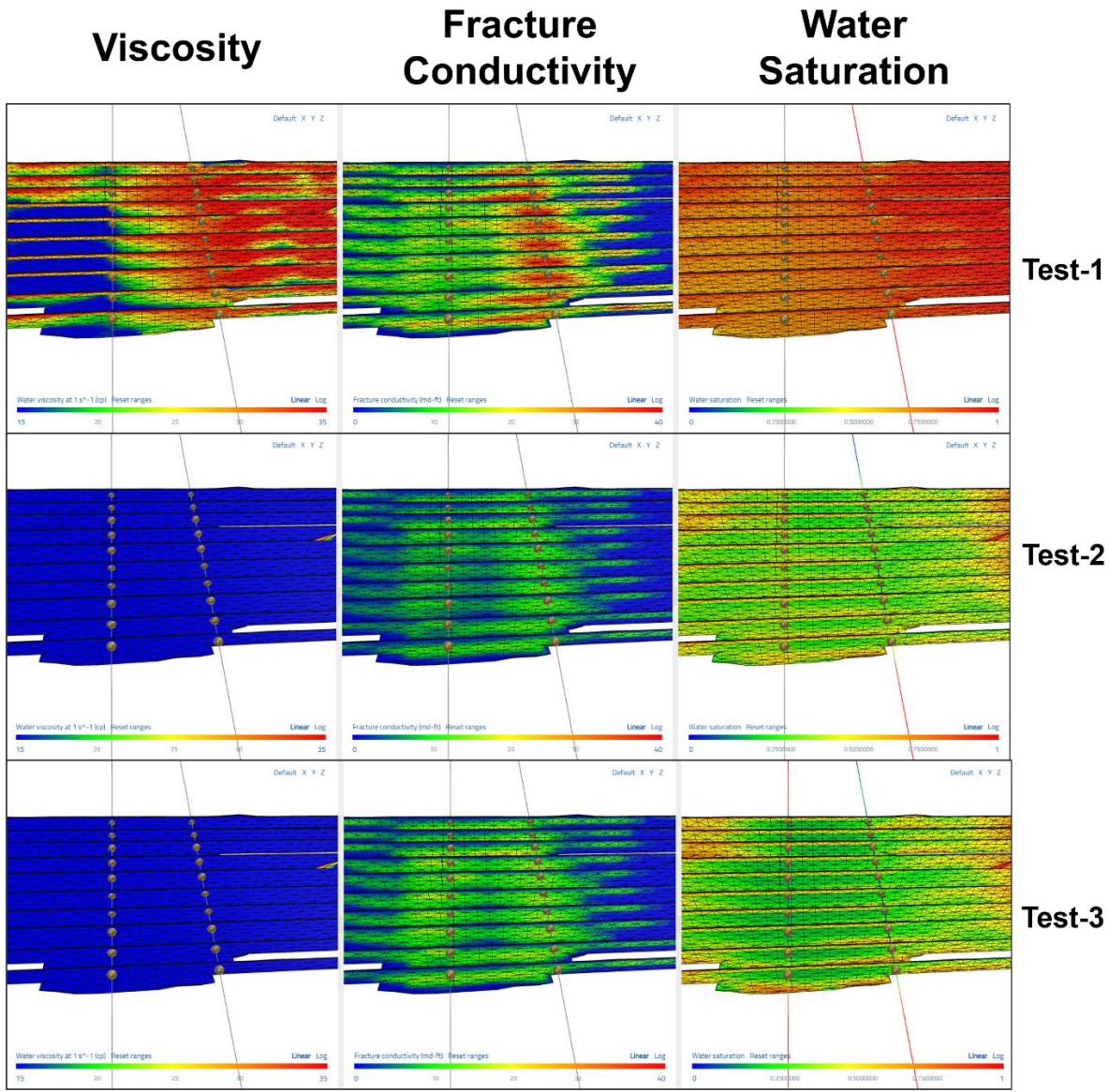

*Figure 22. Water viscosity, fracture conductivity and water saturation in the fractures is shown for the three interference tests. Note the heterogenous distribution of viscosity in the 1st test.*

Because the fracture conductivity is heterogenous both within a single fracture and amongst multiple fractures, the question arises: what conductivity values does the interference test measure? As multiple fractures are analogous to a connection in "parallel", the interference test likely represents the connectivity of the most conductive fracture. Within a fracture the propped elements are analogous to a connection in 'series', and so the interference test measures an approximately harmonic average of the conductivity distribution along the fracture. Figure 23 shows the estimation of fracture conductivity for the three tests outlined in Section 3. For the estimation of the $L_D$ and the $DPI$ plot, the conductivity of Test-2 is used as it is closest to the time when the measurement of the $DPI$ is performed.

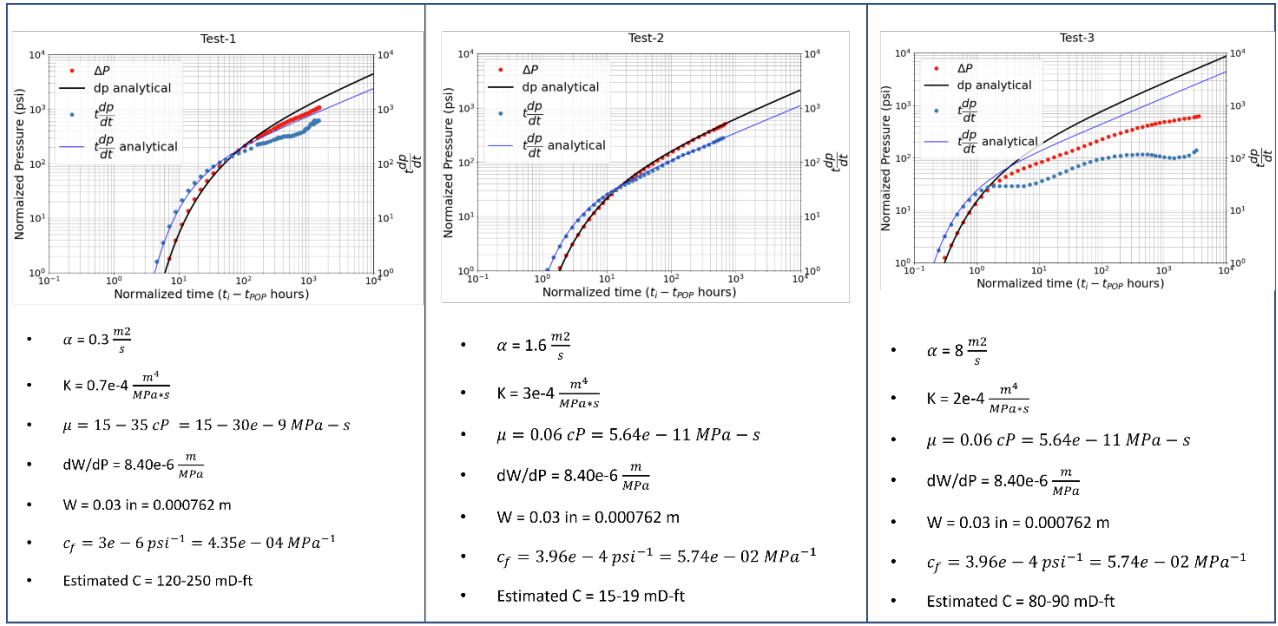

*Figure 23. Fracture conductivity estimates for the three tests in the realistic stage stimulation model using the methodology described in Section 3.*

# Appendix D: Derivation from Radial Flow of the Classical Definition for Dimensionless Fracture Conductivity

Equation 7 expresses the flow rate along a fracture, using Darcy's law:

$$Q_{fracture} = \frac{C_{fracture}}{\mu} H \frac{\Delta p_{fracture}}{L} \tag{D1}$$

Pressure drop from infinite acting radial flow scales as (Horne, 1995):

$$\Delta p_{radial} \propto \frac{Q_{radial}\mu}{kH} \tag{D2}$$

Setting the radial flow rate equal to the fracture flow rate, and solving for the ratio of $\Delta p_{radial}$ and $\Delta p_{fracture}$ yields:

$$\frac{\Delta p_{radial}}{\Delta p_{fracture}} \propto \frac{C_{fracture}}{kL} \equiv F_{CD} \tag{D3}$$

Equation D3 is the classical definition of dimensionless fracture conductivity. Comparison with Equation 12 shows that the scaling is different when considering either linear or radial flow.

Figure 24 show the $F_{CD}$ derived from the classical radial flow assumption plotted against DPI. Because the $F_{CD}$ does not account for changes in fluid and matrix properties appropriately, there is not a clean relationship.

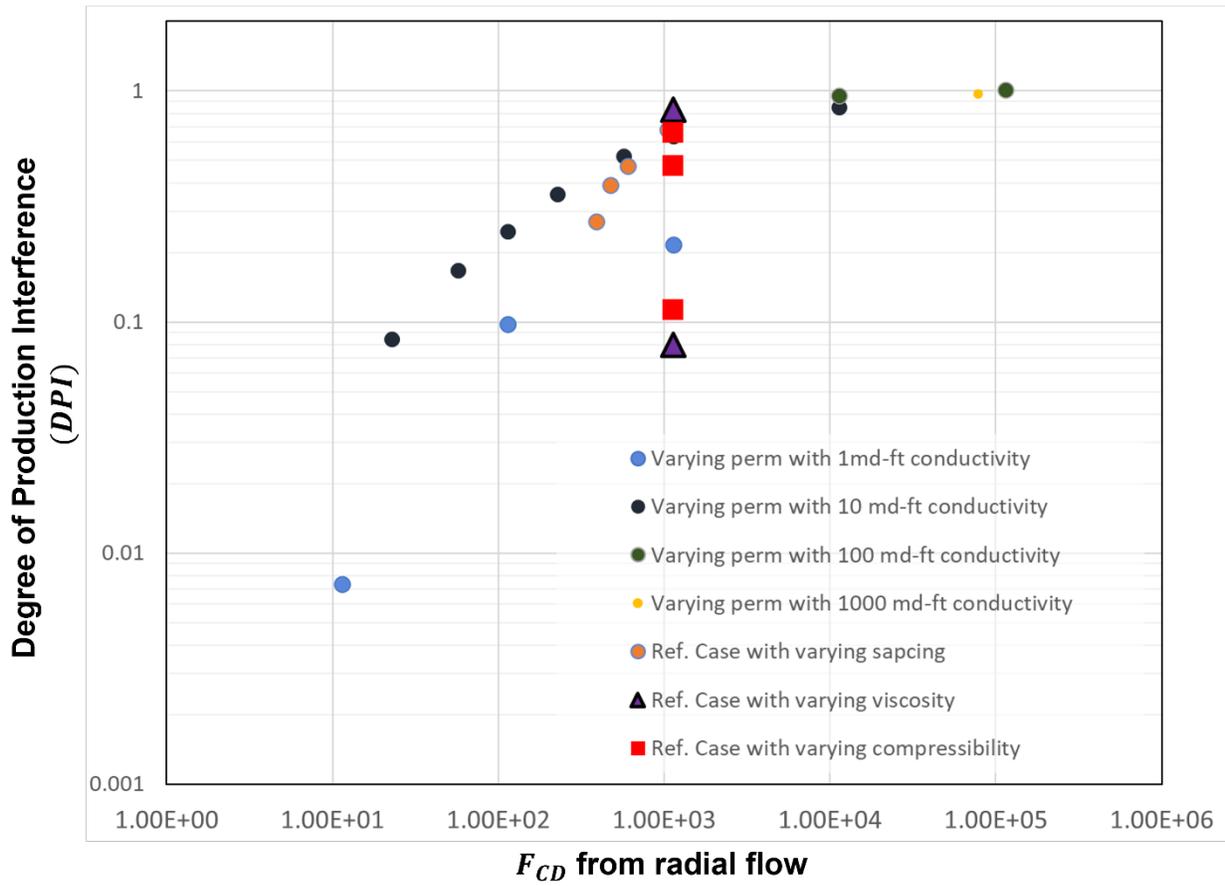

Figure 24. $F_{CD}$ derived from radial flow assumption is plooted against DPI. Note the variation of DPI for points with the same $F_{CD}$ when viscosity and compressibility change.